\documentclass[namedreferences]{solarphysics}

\usepackage[hyperref,optionalrh,showbiblabels]{spr-sola-addons} 
\usepackage{graphicx}        
\usepackage{color}           
\usepackage{amsmath}
\usepackage{appendix}

\defcitealias{Liewer2019}{I}

\newcommand{\gammabeta}{[\gamma(t_i), \beta(t_i)]}

\newcommand{\rsun}{$R_{sun}$}

\begin{document}
\begin{article}
\begin{opening}

\title{Extracting the Heliographic Coordinates of Coronal Rays using Images from  WISPR/{\it Parker Solar Probe}}


\author[addressref=aff1]{\inits{}\fnm{P. C.}~\lnm{Liewer}}
\author[addressref=aff2]{\inits{}\fnm{J.}~\lnm{Qiu}}
\author[addressref=aff2]{\inits{}\fnm{F. }~\lnm{Ark}}
\author[addressref=aff1]{\inits{}\fnm{P.}~\lnm{Penteado}}
\author[addressref=aff3]{\inits{}\fnm{G.}~\lnm{Stenborg}} 
\author[addressref=aff3]{\inits{}\fnm{A.}~\lnm{Vourlidas}}
\author[addressref=aff1]{\inits{}\fnm{J. R.}~\lnm{Hall}}
\author[addressref=aff4]{\inits{}\fnm{P.}~\lnm{Riley}}


\address[id=aff1]{Jet Propulsion Laboratory, California Institute of Technology, Pasadena, CA, 91109, USA (email: Paulett.C.Liewer@jpl.nasa.gov)}
\address[id=aff2]{Department of Physics, Montana State University, Bozeman, MT, 59717, USA}
\address[id=aff3]{Johns Hopkins University Applied Physics Laboratory, Laurel, MD, 20723, USA}
\address[id=aff4]{ Predictive Science Inc., San Diego, CA, 92121, USA}

%
\runningauthor{P.~C. Liewer et al.}

\begin{abstract}

The {\it Wide-field Imager for Solar Probe} (WISPR) onboard {\it Parker Solar Probe} (PSP), observing in white light, has a fixed angular field of view, extending from 13.5$^{\circ}$ to 108$^{\circ}$ from the Sun and approximately 50$^{\circ}$  in the transverse direction. In January 2021, on its seventh orbit, PSP crossed the heliospheric current sheet (HCS) near perihelion at a distance of 20 solar radii.  
At this time, WISPR observed a broad band of highly variable solar wind and multiple coronal rays. For six days around perihelion, PSP was moving with an angular velocity exceeding that of the Sun. During this period, WISPR was able to image coronal rays as PSP approached and then passed under or over them.  We have developed a technique for using the multiple viewpoints of the coronal rays to determine their  location (longitude and latitude) in a heliocentric coordinate system and used the technique to determine the coordinates of three coronal rays. The technique was validated by comparing the results to observations of the coronal rays from {\it Solar and Heliophysics Observatory} (SOHO) / {\it Large Angle and Spectrometric COronagraph} (LASCO)/C3 and {\it Solar Terrestrial Relations Observatory} (STEREO)-A/COR2. 
Comparison of the rays' locations were also made with the HCS predicted by a 3D MHD model. 
In the future, results from this technique can be used to validate dynamic models of the corona. 


\end{abstract}


\end{opening}

\section{Introduction}

The {\it Wide-field Imager for Solar Probe} \citep[WISPR:][]{Vourlidas2016} onboard the {\it Parker Solar Probe} \citep[PSP:][]{Fox2016} 
is returning white-light images of the corona from its unprecedented vantage point inside the orbit of Mercury. 
White light coronagraphs observe sunlight Thomson-scattered by the electrons in the corona. This scattering cross-section has a broad maximum when the Sun-electron-telescope angle is 90$^{\circ}$, a region referred to as the Thomson sphere.
A coronagraph image records the integrated Thomson-scattered signal from all the electrons along the line of  sight and thus is weighted measure of the integrated electron density along the line of sight \citep{Vourlidas2006}. 

The coronal magnetic fields strongly influence the structure of the white-light corona. 
The Sun’s corona received its name from the crown of bright visible rays seen first during solar eclipses. 
In three seminal papers \citep{Wang1997, Wang1998, Wang2000}, it was shown that the SOHO/LASCO coronagraph observations of coronal rays could be reproduced by assuming a thin (3-5$^{\circ}$) uniform sheet of plasma surrounding the  heliospheric current sheet (HCS). This sheet is the coronal counterpart to the heliospheric plasma sheet (HPS). They found that the coronal rays in the images resulted from folds in the warped current sheet, which led to longer paths of integration through the  plasma sheet  along some lines of sight. The plasma was hypothesized to come from the cusps of helmet streamers as a result of reconnection. The  plasma then flows out along open field lines around the HCS. 
Synthetic coronagraph images created from MHD models were also able to reproduce the dominant coronal streamer belt structure seen in LASCO images \citep{Linker1999}. 
\citet{Liewer2001} found that some coronal rays located near the HCS resulted from enhanced plasma outflows associated with active regions. 
Subsequent studies suggested that some features in coronagraph images may not be explained with a single plasma sheet\citep{Saez2005}.  Indeed, \citet{Wang2007}, from an analysis of eclipse images and forward modeling, 
determined that, in addition to coronal rays associated with the HCS, there were rays resulting from pseudo-streamer plasma sheets and also rays associated with bipolar regions in coronal holes (polar plumes). 
\citet{Thernisien2006} found that large density variations (up to 10x) along the sheet were required to reproduce the LASCO observation of a helmet streamer.
\citet{DeForest2018}, using long exposure COR2 observations and sophisticated data analysis, concluded that the corona showed radial structure with high density contrast at all observable scales down to the limits of the instrument, suggesting a continual reconfiguration of the source region.

\citet{Poirier2020} performed the first analysis of the high resolution views of coronal rays from WISPR using images from PSP's  first orbit  ($r_p $= 35.7 $R_{sun}$). They concluded that WISPR “acts like a microscope providing a blown-up view of streamers.” Fine-scale structure was found in the densest part of the streamer rays identified as the solar origin of the heliospheric plasma sheet. They used MHD models and synthetic images to determine the origin of the rays. Some were related to folds in the HCS not discernible from 1 AU and others seems to result from the inherently inhomogeneous distribution of open flux tubes. 
The Thomson sphere diameter is the distance from the telescope to the Sun, so as PSP's perihelion decreases, WISPR's sensitivity to local features  continues to increase\citep{Vourlidas2016}.

 During the seventh orbit, PSP went deeper into the corona 
 ($r_p $= 20.4 $R_{sun}$ at 2021-01-17T17:37), and the inner edge of WISPR's field-of-view, at an elongation of 13.5$^\circ$, could reach as close to the Sun as  5 \rsun. 
 Fortuitously, three closely spaced crossings of the Heliospheric Current Sheet (HCS) occurred 2-3 hours before perihelion, providing views of the streamer belt from within.
 For several days during the period when PSP’s angular velocity was faster than the Sun’s, WISPR observed multiple coronal rays as they approached and then passed over or under the spacecraft. This rapid change in apparent latitude of the rays provides information on their 3D location relative to PSP \citep{Liewer2019}.  
We have developed a technique for using a time sequence of WISPR images of the coronal ray to determine its 3D location in a heliocentric coordinate system under the assumption that the ray has fixed angular coordinates. The technique is a modification of the Tracking and Fitting technique used to determine the trajectory of CMEs from a sequence of WISPR images \citep{Liewer2019, Liewer2020}.	
Using this technique, the coordinates (latitude and longitude) of three coronal rays were determined. The validity of the technique was tested by comparing with simultaneous observations of the corona  and its rays from other white light coronagraphs, {\it Solar and Heliophysics Observatory} (SOHO) / {\it Large Angle and Spectrometric COronagraph} (LASCO)/C3 and  {\it Solar Terrestrial Relations Observatory} (STEREO)-A/COR2.  We compared the locations of the rays to the LASCO/C3 synoptic white light maps for this time period.  More detailed comparisons were made  by using the coordinates of the solution to generate a series of 3D points along the ray and projecting these points onto simultaneous images of the corona from COR2 and LASCO/C3 with different vantage points. 
The goal of this work is to develop and validate this technique and use the determination of individual ray locations  to help validate  3D models of the corona.

The organization of this article is as follows: In Section 2, the technique for determination of the coronal ray coordinates in a heliocentric coordinate system is described. In Section 3, the results of the determination of coordinates of three rays are presented. In Section 4, the results are validated by comparison to synoptic white light images from LASCO/C3, to images of the coronal rays from LASCO/C3 and COR2A, and to synoptic results from the MHD model CORHEL. Section 5 contains a summary and discussion of the results.


\section{Tracking and Fitting Technique}
\begin{figure}    
   \centerline{\includegraphics[width=1.0\textwidth,clip=]{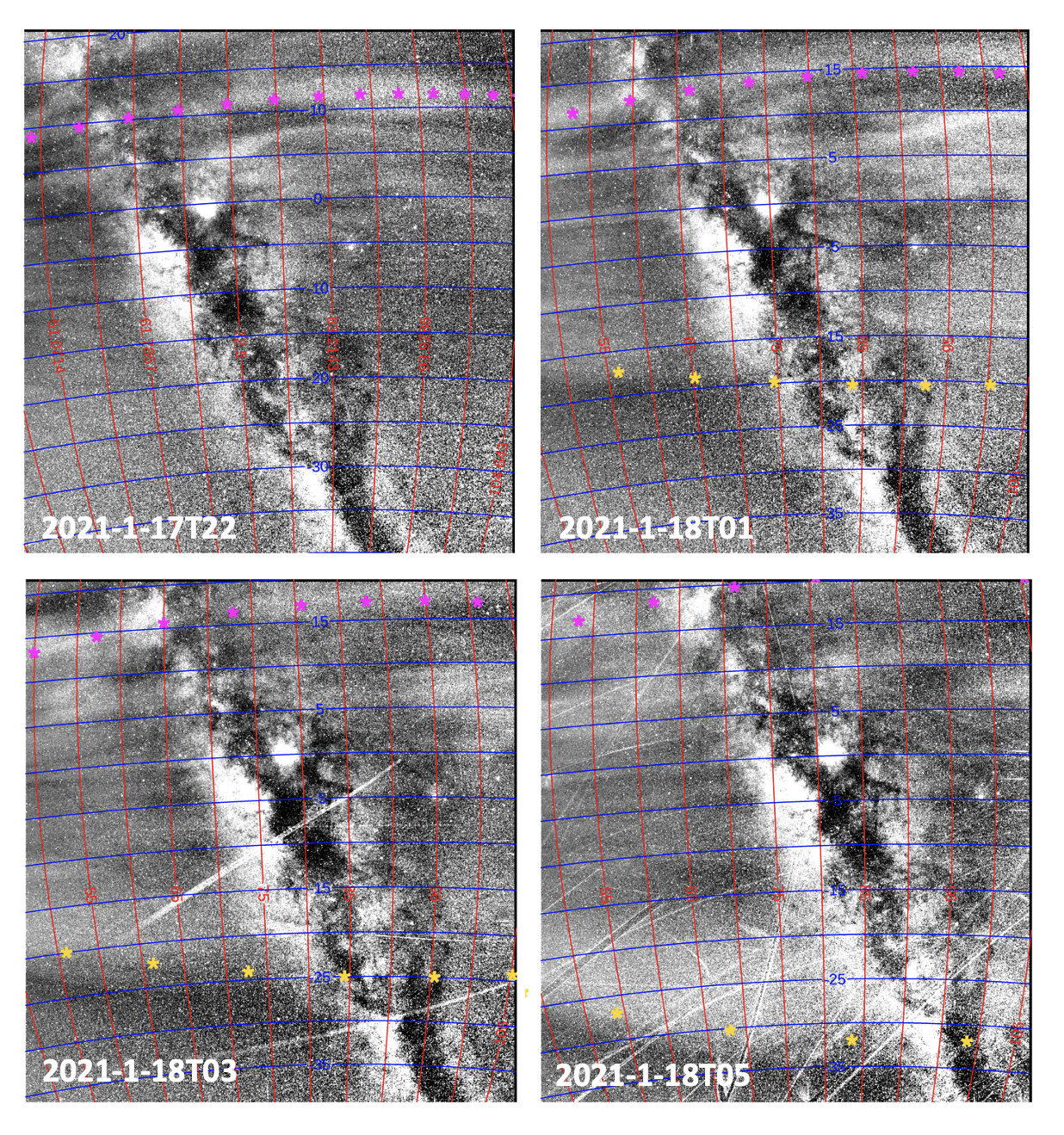}
              }

	\caption{
WIS-O images at four times showing the apparent changes in latitude of two coronal rays as the separation between the rays and the spacecraft decreases. The upper ray, WIS-O High Ray 1(magenta symbols) can be seen in all four images as it moves upward and then leaves through the upper boundary. The lower ray, WIS-O Low Ray (yellow symbols) can be seen in the last three images as it moves downward in the FOV. The Milky Way is evident in all four images, slowly moving across the  FOV. The streaks are due to sunlight reflecting off debris created by dust impacts on the spacecraft.
	}
   \label{fig:2rays}
   \end{figure}

During each orbit, PSP’s angular velocity exceeds that of the Sun for some number of days around perihelion. Thus, WISPR is imaging quasi-stationary coronal features as it flies through them. The technique for determining 3-D location of a coronal ray from a sequence of WISPR images is based on the fact that the {\it apparent latitude} of a quasi-stationary feature increases as it is approached by PSP [see, e.g. \citet{Liewer2019}]. 
This increasing apparent latitude is illustrated in Figure~\ref{fig:2rays},
which shows a time sequence of images as WISPR approaches two coronal rays. The four images are from the outer
telescope, WISPR-O (50.5$^\circ$ -  108.5$^\circ$ elongation from Sun center). They were taken near perihelion ($r_p $= 20.4 $R_{sun}$),  covering the period 2021-01-17T22  to 2021-01-18T05. The images shown here have been processed using a technique, referred to as LW processing, which is designed to enhance the changes in the images from one to the next.
The LW technique exploits the time domain to create a background model for each image by estimating the baseline brightness at the 5-percentile level on each pixel at each time instance. The 5-percentile basal level is estimated considering a running window of a certain time length. The choice of the window length will affect how much small-scale detail is enhanced in the background-subtracted images (the smaller the window size, the finer the structure revealed, akin to a high pass filter). For the present work, we used a window size equivalent to the time elapsed during nine images. The technique is described in detail in the Appendix to \citet[][in press]{Howard2022}. 
Thus, this processing captures the apparent increase in latitude of a coronal ray as the distance between the ray and PSP decreases and the apparent latitude changes. 

In Figure~\ref{fig:2rays}, the ray  marked with a series of magenta symbols can be seen in all four images as it moves up and then leaves the WISPR-O field-of-view (FOV) through the upper boundary, indicating that PSP passed under it. The second ray, marked with yellow symbols, can be seen only in the last three images as it moves toward the lower boundary of the FOV. Thus, the spacecraft passed between these two rays. The rate of increase of the apparent latitude depends on the distance from the spacecraft and this, together with the assumption that the coronal ray has fixed angular coordinates in a heliocentric frame, is sufficient information to extract the angular coordinates of the ray. Said another way, the assumption of constant angular coordinates allows one to use the changing position of the coronal ray in the WISPR images to relate its pixel coordinates in the image to its assumed-constant angular coordinates in a 3D heliocentric frame.  The technique described here is a modification of the Tracking and Fitting technique developed to determine the trajectories of solar ejecta [\cite{Liewer2019}, hereafter Article I;  \cite{Liewer2020}] and many of the details are the same. 
 
\begin{figure}    
\centerline{\includegraphics[width=0.8\textwidth,clip=]{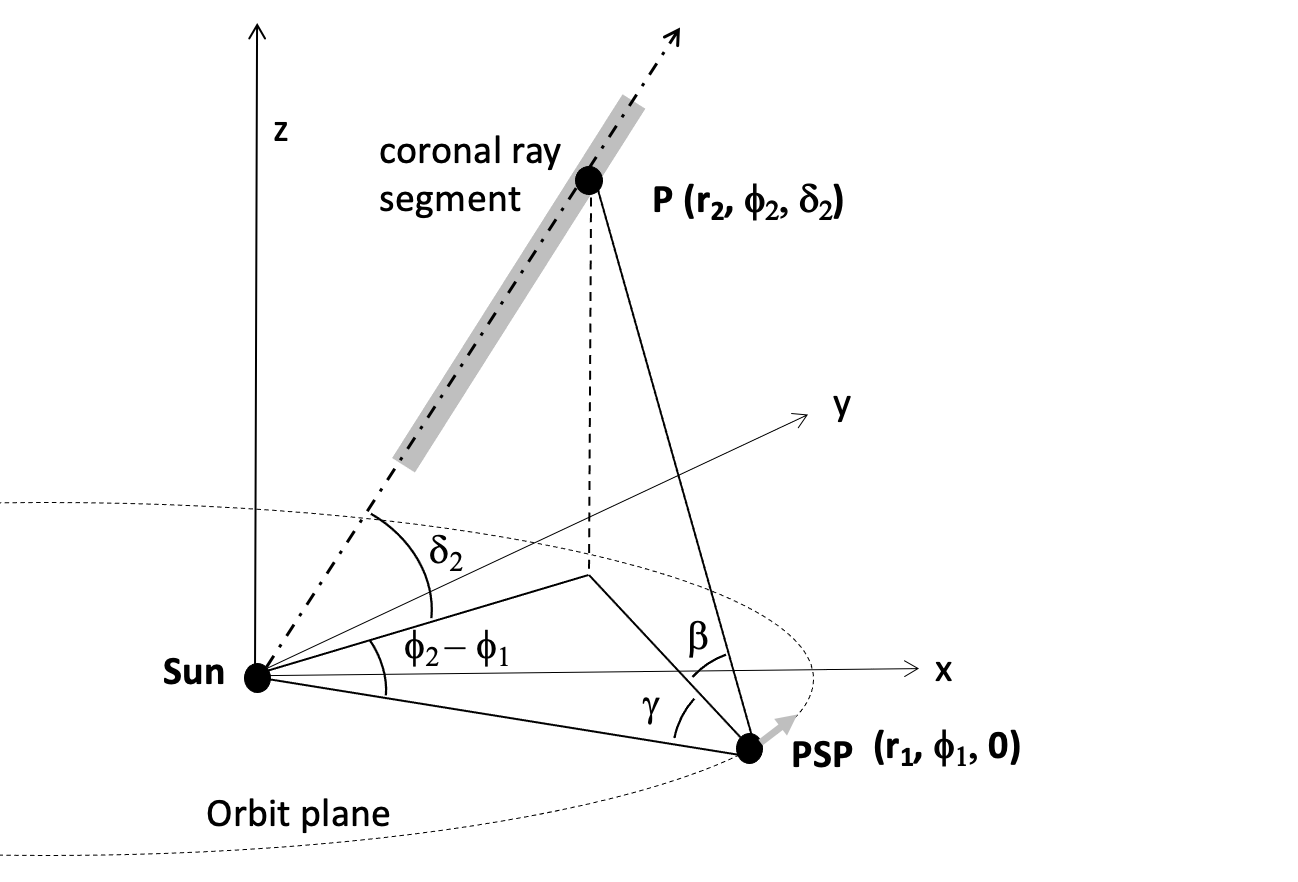}
              }
	\caption{Geometry relating the coordinates of a point P along a coronal ray ($r_2,\phi_2,\delta_2$) in the HCI coordinate frame to the two angles $\gamma,\beta$ defining the unique LOS from PSP to the point P under the assumption that PSP's orbit lies in the solar equatorial plane. (Adapted from Article \citetalias{Liewer2019}.)}
   \label{fig:geom}
   \end{figure}

\subsection{Coordinate Systems and the Equations for Fitting}

The geometry relating the view of the coronal ray from WISPR to its coordinates in the Heliocentric Inertial (HCI) frame is shown in Figure~\ref{fig:geom}. This is similar to Figure~11 in  Article \citetalias{Liewer2019}.
Here, for clarity of presentation, we first assume that PSP orbits in the solar equatorial plane. This assumption is relaxed in the actual fitting procedure described below. 
The HCI coordinates are $[r, \phi, \delta]$, where $r$ is the distance to the Sun, $\phi$ is the 
angle (longitude) in the solar equatorial plane (the $x$-$y$ plane), and $\delta$ is the angle  (latitude) out of this plane. 
In this frame, the known (from the ephemeris) time-dependent coordinates of PSP (dotted line) are $[r_1, \phi_1, 0]$.
The  segment of the ray seen by WISPR is shown as the thick gray line falling along a radial line in the HCI frame. The coordinates of points along this segment
 are $[r_2, \phi_2, \delta_2]$. 
Each pixel in the WISPR image defines a unique line of sight (LOS) from the spacecraft, specified by two angles, $\gamma$ and $\beta$, where
$\gamma$ is the angle in the orbit plane measured with respect to the Sun-PSP line and $\beta$ is the angle out of the orbit plane. In this reference frame, which we refer to as the PSP orbit frame, the Sun is at  [$\gamma, \beta] = [0,0]$. The position of a coronal ray at time $t$ will thus be described by a set of angles  $\gamma(t)$ and $\beta(t)$ measured along the ray.

Using basic trigonometry, we obtained the equation relating the HCI coordinates of a point along the ray to its coordinates in the observer-based PSP orbit frame. 
\begin{flalign}
& \frac{\tan\beta (t)}{\sin\gamma (t)} = \frac{\tan\delta_2}{\sin[\phi_2 - \phi_1 (t)]}, \label{eq:1}   
\end{flalign}

Equation ~\ref{eq:1} relates the feature's angles in the HCI frame to its angles $\gamma$ and $\beta$ referenced to the location of the PSP spacecraft; 
there is no dependence on either $r_1$ or $r_2$. This is the same equation as the first of two equations used for trajectory determination in Article \citetalias{Liewer2019}. We now make the assumption that all points along the coronal ray segment have the same HCI angular coordinates $\phi_2$ and $\delta_2$, e.g., the coronal ray segment falls along a radial line in inertial space. By obtaining a set of $\gammabeta$ measurements from a time sequence of WISPR images, Equation ~\ref{eq:1}  can be solved to determine the unknown, constant angles $\phi_2$ and $\delta_2$.  This is the basis for the technique for determining the ray's location.  As PSP approaches the ray, the denominator $\sin(\phi_2-\phi_1)$ approaches $0$, leading to a rapid increase in $\beta$.

Note that, in principle, for constant {$\phi_2$ and $\delta_2$} along the ray, the ratio $\mathcal{R} \equiv {\tan\beta/\sin\gamma}$ should be constant for every point along the ray measured from a given image at time $t$. In practice, we obtained several $[\gamma, \beta]$ pairs for multiple points identified along the same ray segment in the image to compute the mean ratio $\langle\mathcal{R}\rangle$ for this time, and its standard deviation $\Delta \mathcal{R}$ as an approximation of the measurement uncertainty in $\langle\mathcal{R}\rangle$. This ratio $\langle\mathcal{R}\rangle$ changes when PSP's position $\phi_1$ changes with time. The measurements of $\langle\mathcal{R}\rangle$ and $\Delta \mathcal{R}$ are made at multiple times over a period of about five hours as PSP rapidly moves on its orbit with $\phi_1$ varying for 5$^{\circ}$ - 10$^{\circ}$.  
A least-squares fit of the measured $\langle\mathcal{R}\rangle$ to Equation~\ref{eq:1} would return the HCI coordinates $[\phi_2, \delta_2]$ of the ray, subject to the above approximations and limitations.

Recall that Equation~\ref{eq:1} is derived assuming that the spacecraft orbit lies in the solar equatorial plane. This is a reasonable approximation since PSP's orbit is close to Venus' orbital plane, which is inclined by about 4$^{\circ}$ to the solar equatorial plane. The inclination of PSP's orbit relative to the solar equatorial plane changes each time PSP uses Venus' gravity to reduce the perihelion. Equation~\ref{eq:1} is only used to give an initial guess for $[\phi_2, \delta_2]$. The actual equation used to fit the data 
is given in the Appendix (Equation~\ref{eq:correct}), which includes a first order correction for the small inclination angle $\epsilon$ of the PSP orbit plane with respect to the solar equatorial plane. The correction involves a coordinate transformation from a heliocentric coordinate system defined with PSP's orbital plane to the HCI coordinate system, which is identical to that made in the Tracking and Fitting technique in Article \citetalias{Liewer2019}; the geometry relating the PSP orbit frame to the HCI frame is shown in Article \citetalias{Liewer2019}, Figure 2. The PSP orbit frame is defined by PSP's velocity vector and the vector from the Sun to PSP's current location.
Note that for $\delta_2 \approx 0$, Equation~\ref{eq:1} becomes trivial and cannot be used to find $\phi_2$; this is overcome with the corrected Equation~\ref{eq:correct} that includes the inclination of PSP's orbit.  This is further discussed in the Appendix to this article.

Employing the corrected equation (Equation~\ref{eq:correct}), we apply a least-squares curve-fitting algorithm to determine a feature's coordinates in the HCI frame from its positions tracked in WISPR images, given a set of $\langle\mathcal{R}\rangle$ measurements and their uncertainties $\Delta \mathcal{R}$ in the image sequence.

The same technique can also be applied when a different assumption is made about the nature of the coronal ray.  Instead of assuming the ray segment lies along a radial line in the HCI system, we make the assumption that the ray segment is rigidly rotating with the Sun. This is equivalent to the assumption that the ray segment lies along a radial line in the Carrington coordinate frame. In fact, the same data set of $[\gamma,\beta]$ pairs can be used to obtain the solution in the Carrington frame as well as in the HCI frame.  

The details of the fitting procedure 
are described in the Appendix. The fitting program returns the angular coordinates and their uncertainties in either HCI or Carrington frame, and also the reduced $\chi^2$, defined in the Appendix (Equation~\ref{eq:chisq}), as an indicator of the overall goodness of fit to the data set. 

Both assumptions -- that the ray segment is radial in an inertial frame and that the ray segment is in rigid rotation with the Sun -- are undoubtedly approximations to the true geometry of the coronal ray. The ray may have curvature on a larger spatial scale. The ray may have a significant lateral extent as well, as would be the case if it resulted from a LOS passing through a folded plasma sheet \citep[see, e.g.,][]{Thernisien2006}.

 \subsection{Obtaining the data set from the images}

\begin{figure}    
   \centerline{\includegraphics[width=1.0\textwidth,clip=]{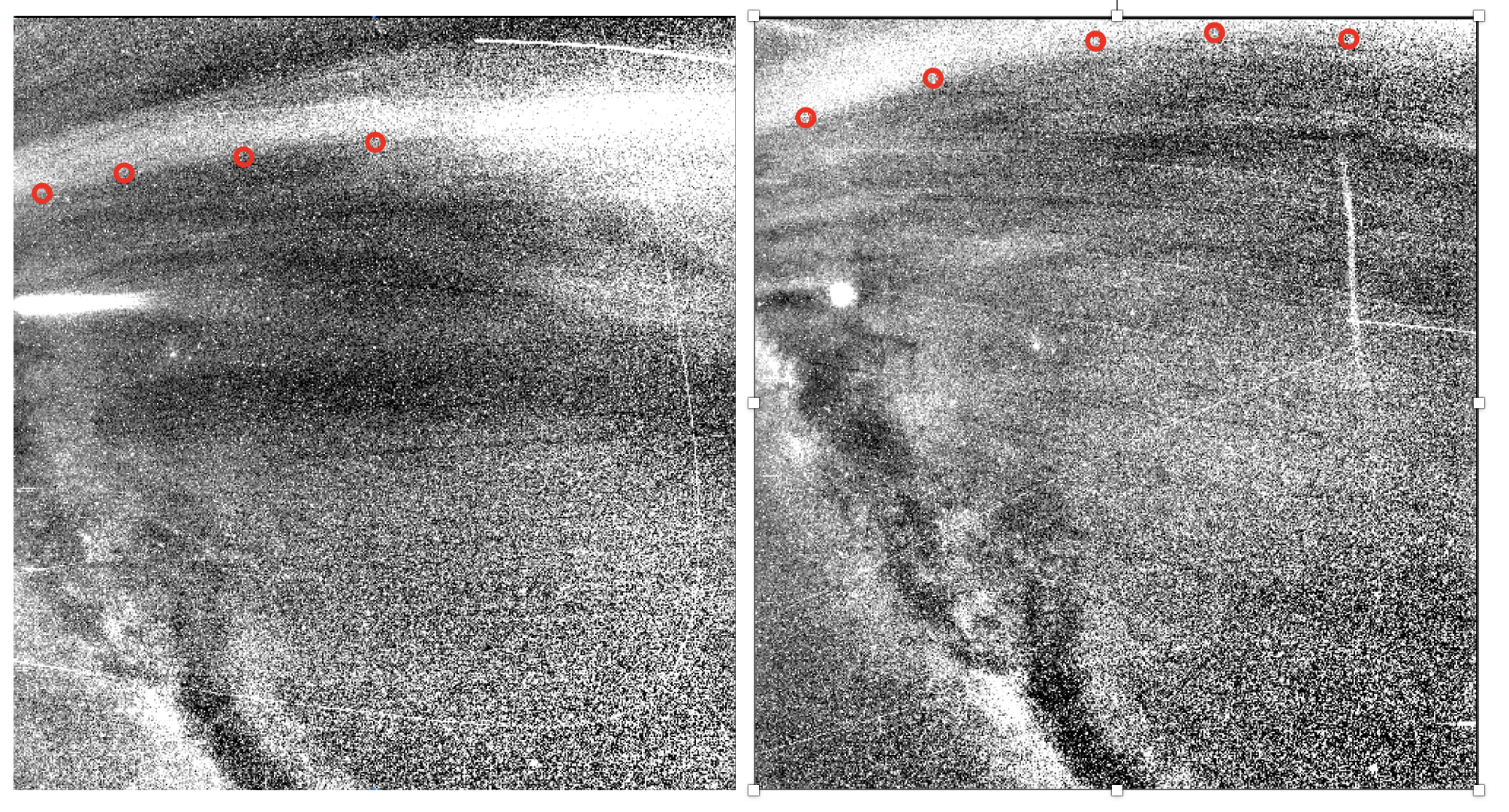}
              }
	\caption{
WISPR-O images showing sample user-selected data points for tracking WIS-O High Ray 2 at two times, the start and end times given in Table 1. The tracking software places red circles around points selected manually by the user and saves the coordinates to a file along with the other information needed to perform the fit. Here, the thickness of the circle has been increased for visibility. 
	}
   \label{fig:sample_data}
   \end{figure}

To determine the location of a coronal ray segment, we start by obtaining the image coordinates of a set of points  along the coronal ray  for a sequence of images, typically with a one hour cadence. This is done manually by tracing and tracking the ray using a cursor. The manual tracing results in sets of pixel coordinates; the pixel coordinates of the points are converted into the observer-based angles  $(\gamma, \beta)$  in the PSP orbit frame, defined above and in Article \citetalias{Liewer2019}. To obtain an estimate of the uncertainty in our data, we select data points at two or more positions along the ray for each image (each time) in the sequence, as discussed above, and compute the mean of the ratio $\mathcal{R} \equiv \tan\beta/\sin\gamma$ and its deviation $\Delta \mathcal{R}$. The time sequence of the mean ratio $\langle\mathcal{R}\rangle$ and its uncertainty $\Delta\mathcal{R}$, along with the coordinates of PSP, are fed into the fitting program to determine the constant angular coordinates  $\phi_2$ and $\delta_2$ and their uncertainties in the HCI and/or Carrington frames of reference. The image sequences used in the tracking cover four to five hours. 
Figure~\ref{fig:sample_data} shows sample data points for two of the images used in tracking one of the rays discussed below (WIS-O High Ray 2). These are WISPR-O images at the start (2021-01-17T09:05) and end (2021-01-17T14:05) of the sequence of images used in the tracking. The red circles mark the location of the pixel coordinates selected by the cursor. Here, the circles have been made thicker to improve visibility. Note that several locations along the ray have been selected in both images to calculate the deviation in $\Delta \mathcal{R}$ for that image (time).
Figure~\ref{fig:2rays} showed several of the processed images used to track the two coronal rays marked in the images with colored symbols. The points shown in that figure are not the data points used to obtain the solution; they are points generated from the solution which have been projected back onto the image to check the solution; this is discussed further below. 

Note in Figure~\ref{fig:sample_data} that the points have been selected along the lower edge of the ray because this is a  well-defined feature.  For a ray to be tracked, it must be clearly visible and well defined for a span of four or more hours and have a significant change in apparent latitude. Many faint rays are seen that have significant changes in apparent latitude, but are too faint or too short lived to track. 
For the rays that can be tracked,  the fitting solutions are rather insensitive to the exact placement of the data points.  We know this from re-tracking the same feature and repeating the fitting. 
When multiple tracking and fittings were done, the solution presented here is the one with the best fit as measured by the reduced $\chi^2$. For the rays that can be tracked, the errors in the tracking and fitting are smaller than the uncertainty introduced by the underlying assumptions of the technique, discussed in the next section.

\section{Results}

\begin{figure}    
\centerline{\includegraphics[width=14cm,clip=]{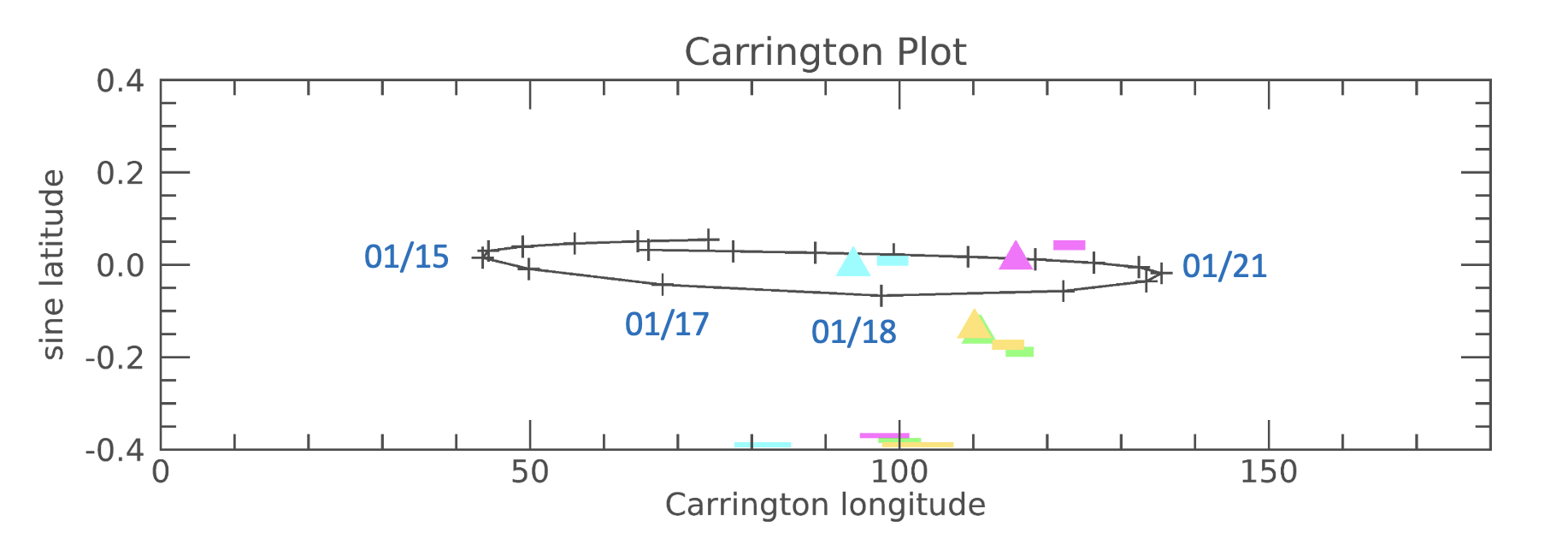}
              }
	\caption{Carrington plot of the coordinates of both the HCI- and Carrington-frame solutions for the four coronal rays whose locations were determined. Also shown is the PSP orbit for ten days around perihelion. The Carrington-frame solutions are shown as triangles; the color code of each of the four rays is given in Table I. The HCI-frame solutions are shown as the short color-coded lines connecting the Carrington coordinates of the HCI solution at the start an end time of the tracking for that feature. Also indicated just above the x-axis are thin color-coded lines covering the range of Carrington longitudes that PSP traversed during the time that feature was tracked.
	}
   \label{fig:psporbit}
   \end{figure}

This section presents the result for the determination of the coordinates of three coronal rays in both the HCI and Carrington coordinate frames. One of the rays spanned the entire WISPR FOV; it  was tracked independently in images from the inner and outer telescopes.  The four independent solutions are presented below. The results are summarized in Table~\ref{table:summary} and in Figure~\ref{fig:psporbit}.

The columns of Table~\ref{table:summary} are labeled with a name for the ray. The first row gives  the color used to plot that feature’s solution in Figure~\ref{fig:psporbit} and in many of the images below. The next two rows of the table give the times for the images used at the start and end of tracking the rays.  The following three rows give the longitude, latitude and  their uncertainties from the fit, and then the overall goodness-of-fit indicated by the reduced $\chi^2$ (Equation~\ref{eq:chisq} in the Appendix) for the Carrington (CAR)  frame solution, followed by a row giving the approximate radial  extent of the segment  traced. The next three rows contain the same information for the HCI-frame solution. The final rows give another estimate of the error in the determination of a ray's location, calculated from the mean difference between the HCI- and Carrington-frame solutions, defined below. 

Figure~\ref{fig:psporbit} further summarizes the results. Both the Carrington and HCI solutions for the four ray tracking data sets are shown on a Carrington plot of the PSP trajectory covering 20 days around the perihelion at 2021-01-17T17:37. The lower loop of the orbit occurs when PSP’s angular velocity exceeds that of the Sun; all of the coordinate determinations were made during this period 
near the HCS crossings, which occurred a few hours before perihelion. The Carrington solutions are plotted as color-coded triangles using the colors assigned in Table I; this same color coding of the four features is used throughout. 
To plot the HCI solution on a Carrington plot, we have computed the Carrington longitudes at the start and end times of the tracking and connected these two angles with a line, leading to a short range of longitudes for the HCI solutions; the latitude is unchanged in the coordinate transformation. The short line connecting the angles is plotted in the same color as the Carrington solution. 
Just above the x-axis of the plot, the range of PSP’s Carrington longitudes traversed during the tracking of each feature is also shown, plotted in the same color as the feature, but using a thinner line. For example, the Carrington longitude of  WISPR-O High Ray 1 (115.7$^\circ$, from Table~\ref{table:summary})  is plotted as a magenta triangle; the HCI solution Carrington range is the nearby short magenta line; and the range of PSP's Carrington longitudes indicated by the thin magenta line just above the x-axis was 95$^\circ$-101$^\circ$.  We next discuss each of the four features in detail.

\subsection{WIS-O High Ray 1}
The angular coordinates for the first feature, WIS-O High Ray 1, determined by the Tracking and Fitting procedure for both Carrington and HCI frames, are given in Table I. This ray segment can be seen in Figure~\ref{fig:2rays}, indicated with a series of magenta symbols, the same color used to plot this feature coordinates in Figure~\ref{fig:psporbit}  and throughout.  Because the ray is assumed radial, once the angular coordinates are determined, we can generate 3D points along the ray --  radial in the HCI frame for the HCI solution and radial in the Carrington frame for the Carrington solution. The magenta symbols (asterisks) along WIS-O High Ray 1 in Figure~\ref{fig:2rays} were created in this way from the HCI solution.  We next projected these 3D points back onto the WISPR images using software developed to project any set of 3D HCI coordinate points onto an image from any telescope using the information in the image’s FITS header. In Figure~\ref{fig:2rays}, the same set of HCI points created from the HCI solution for WIS-O High Ray 1 has been projected onto all four images.  

One reason for making these projections, such as those in Figure~\ref{fig:2rays}, is to validate the solution obtained by tracking and fitting; the points should fall on the feature that was tracked and it can be seen that they do. The second use of the projections is to determine the radial range of the ray segment seen in the WISPR images, which is not determined either directly from the fitting solution or from the observations. In Figure~\ref{fig:2rays}, the magenta points seen in the first image start at 16 \rsun\  and extends to 27 \rsun\  with a fixed spacing of 1 \rsun.   
As PSP approaches the ray, the radial extent of the segment seen in the FOV shrinks and, thus, the later images have fewer points and the separation of the points appears to increase. From the projection at the start of tracking, we determine the radial extent of the ray segment  observed by WISPR-O to be 16-27 \rsun\  for the HCI solution ({\it cf} Table~\ref{table:summary}).

 \begin{figure}    
\centerline{\includegraphics[width=1.0\textwidth,clip=]{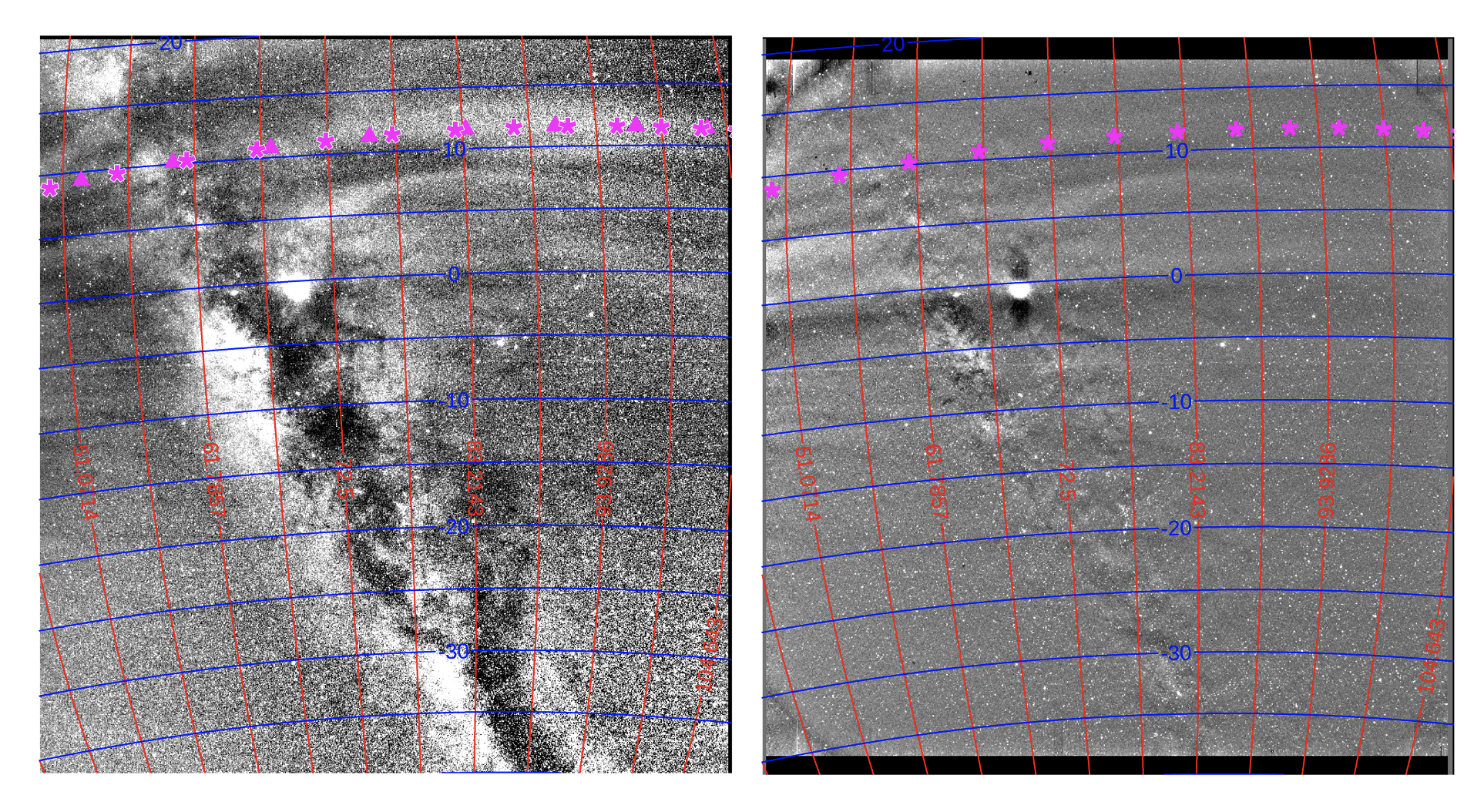}
              }
	\caption{ Left: Projection of the two solutions for the WIS-O High Ray 1 onto one of the LW images used in tracking. The HCI solution is shown as asterisks and the Carrington solutions as triangles. The time of the image is 2021-01-17T22. Right: Projection of the HCI solution onto an L3-processed image at the same time. The bright object is Venus. The Milkway can be seen in both images.
	}
   \label{fig:wisohigh1}
   \end{figure}

We also projected the Carrington solution back onto the images used in tracking. Since the projection software requires the coordinates be supplied in the HCI frame, the Carrington solution must first be converted to HCI coordinates for the time of the image to be used in the projection.  Figure~\ref{fig:wisohigh1} compares the projection of the Carrington solution (triangles) and the HCI solution (asterisks) on 2021-01-17T22.  For the  Carrington-frame solution, the points start at 17 \rsun.  It is evident in the figure that the Carrington solution projects onto the feature just as well as the HCI solution. 
The  $\chi^2$’s for the two solutions are also very close, as shown in Table 1. Thus, we cannot determine which approximation is better – a radial ray rigidly rotating with the Sun or a radial ray fixed in inertial space.  

\begin{figure}    
\centerline{\includegraphics[width=10cm,clip=]{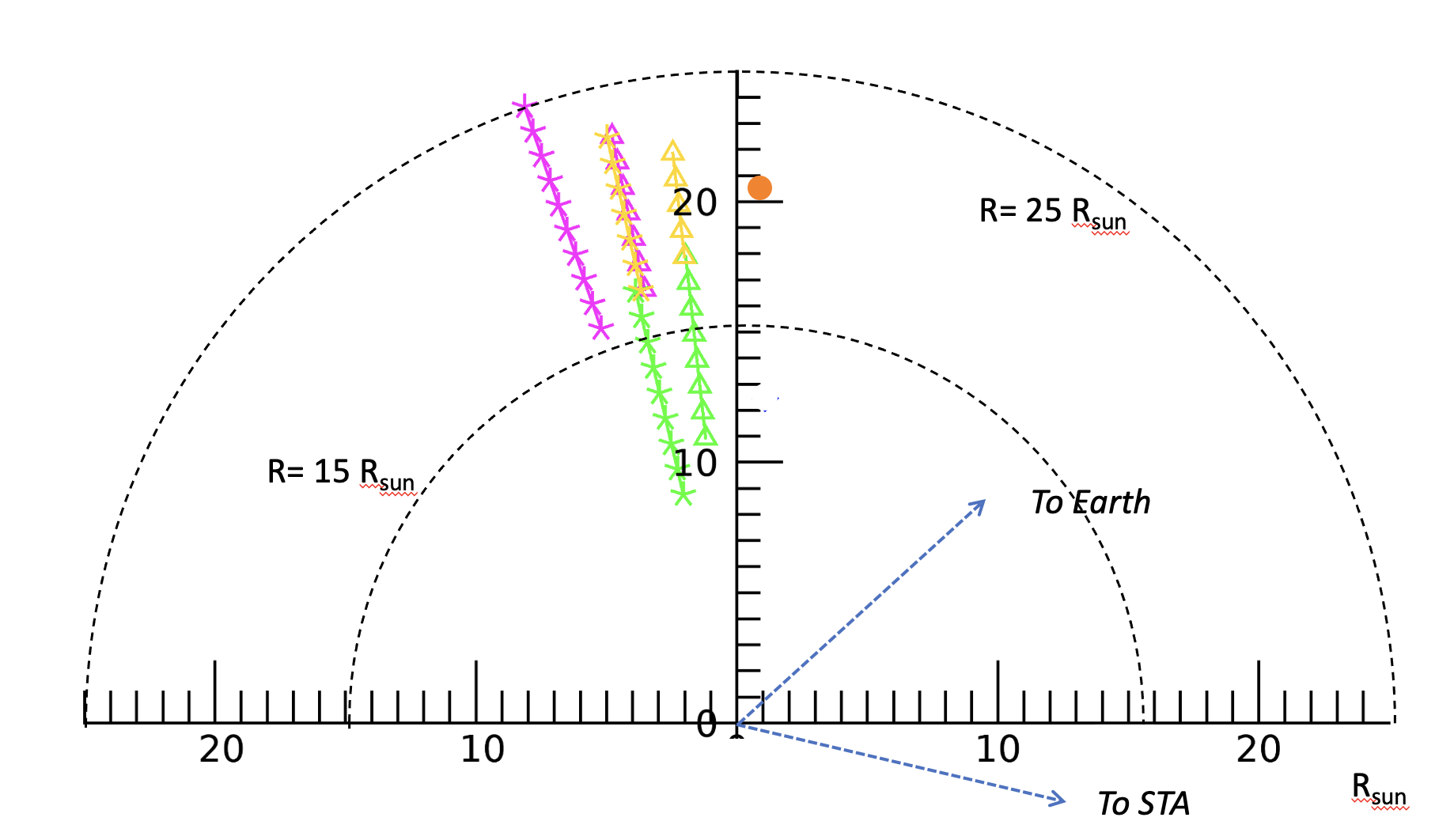}
              }
	\caption{ Polar plot in HCI frame showing both the HCI and Carrington frame solutions for the location of three of the coronal rays tracked.  HCI longitude is measured counter-clockwise from the +x axis, so $90^\circ$ is straight up from the Sun. Both solutions for a  feature are plotted in the color assigned in Table 1. The HCI solution is plotted using asterisks and the Carrington solution plotted using triangles. The Carrington solutions have been converted to HCI angles for the date of 2021-01-18T01, a time when these three features were all visible to WISPR (High Ray 2 was not visible at this time). {\bf The location of PSP at this time is shown as an orange dot.} The directions to Earth (at $42^\circ$ HCI longitude) and to STEREO-A (at HCI $-14^\circ$ longitude) are indicated with blue arrows.} 
   \label{fig:polarplot} 
   \end{figure}

Note that the $\chi^2$ is a measure of how well the data points fit the analytic expression (Eq. 2) and does not reflect other uncertainties introduced by the changing LOS during the tracking or by the assumption of our technique that the angles are constant along the ray segment. As discussed above, both assumptions are  approximations. Instead, we use the the difference in the two solutions to estimate an error in the location of the coronal ray.  We take the difference between the HCI solution longitude and the Carrington solution longitude converted to HCI coordinates at two times: the start and end of tracking, and define the uncertainty in the longitude as the average of these two differences.    Using this estimate for WIS-O High Ray 1, the uncertainty in longitude is  7${^\circ}$. 
Since the latitude of the Carrington solutions is unchanged in the conversion to the HCI frame, the error in the latitude is simply taken to be the difference in the latitudes of the HCI and Carrington solutions, which is 2${^\circ}$ for this ray. This is the way the errors in the location of all the rays in the last two rows of Table 1 were calculated. Note that the errors computed this way are significantly larger than the uncertainties in the angles from the fitting.  From Table 1, it can be seen that the uncertainties in the longitudes from the fitting are on the order of $1^\circ - 2^\circ$ whereas the errors calculated from the difference between the HCI and Carrington solutions were $5^\circ - 7^\circ$.
One might expect a bias in the two solution assumptions towards finding the ray closer to PSP in the Carrington frame than in the HCI frame to make up for the slower approach to the ray in the Carrington frame, since a slower approach lowers the rate of change in the apparent latitude. This is indeed what we observe. The HCI solution ray is always further from PSP than the Carrington solution ray by an amount essentially equal to the error in the longitudes in Table 1. This bias can also be seen in Figure~\ref{fig:polarplot}, discussed below.

The right image in Figure~\ref{fig:wisohigh1} shows the same HCI solution points projected onto a Level 3(L3) processed WISPR-O image at the same time. The L3-processed images, which are the publicly released data product, preserve the calibrated brightness of the images and thus better represent the true density enhancements. The tracked coronal ray is visible in this processed image, but just barely.  There are brighter rays seen lower, closer to the plane of Venus (the bright spot) and, thus, closer to the PSP orbit plane. 
However, these are not as visible in the LW processed image, which records changes from image to image. Evidently, these rays are not changing much in time. Based on their brightness, this is most likely because they lie close to the PSP orbit plane. 
Features that lie in the PSP orbit plane have no change in apparent latitude as approached by PSP.


The  polar plot in Figure~\ref{fig:polarplot}, in the HCI frame of reference,  provides another comparison of the Carrington and HCI solutions for the WIS-O High Ray 1 on 2021-01-18T01. The radial extent of the tracked segments is also shown. Again, WIS-O High Ray 1 is shown as magenta  asterisks  for the HCI solution and magenta triangles for the Carrington solution.  From this plot, the uncertainty in the longitude, calculated above at 7${^\circ}$, can also be estimated by eye. Both solutions for the other two coronal ray segments that were visible at this time, WIS-I Low and WIS-O Low, are also shown in their assigned colors.  Here, HCI longitude is measured counter-clockwise from the +x axis, so $90^\circ$ is straight up from the Sun. The location of PSP at this time is shown as an orange dot. The directions to Earth, at  $42^\circ$ longitude, and to STEREO-A, at  $-14^\circ$ longitude, are indicated with blue arrows.  Thus the coronal rays should be visible in images from COR2A and LASCO, albeit with different lines of sight.

\subsection {WIS-O High Ray 2}
The angular coordinates for the second feature, WIS-O High Ray 2, as determined by the Tracking and Fitting procedure for both Carrington and HCI frames, are given in Table I, along with the related information for this feature. The angular coordinates were plotted in Figure~\ref{fig:psporbit} in cyan.  This feature was seen earlier in the orbit as indicated in Table 1 and as is evident from the range of PSP longitudes during the tracking, shown in Figure~\ref{fig:psporbit} as the cyan bar on the x-axis. This ray in not included in the  polar plot in 
Figure~\ref{fig:polarplot}, which shows positions on 2021-01-18T01, because this ray was not seen at this time.

\begin{figure}    
\centerline{\includegraphics[width=1.0\textwidth,clip=]{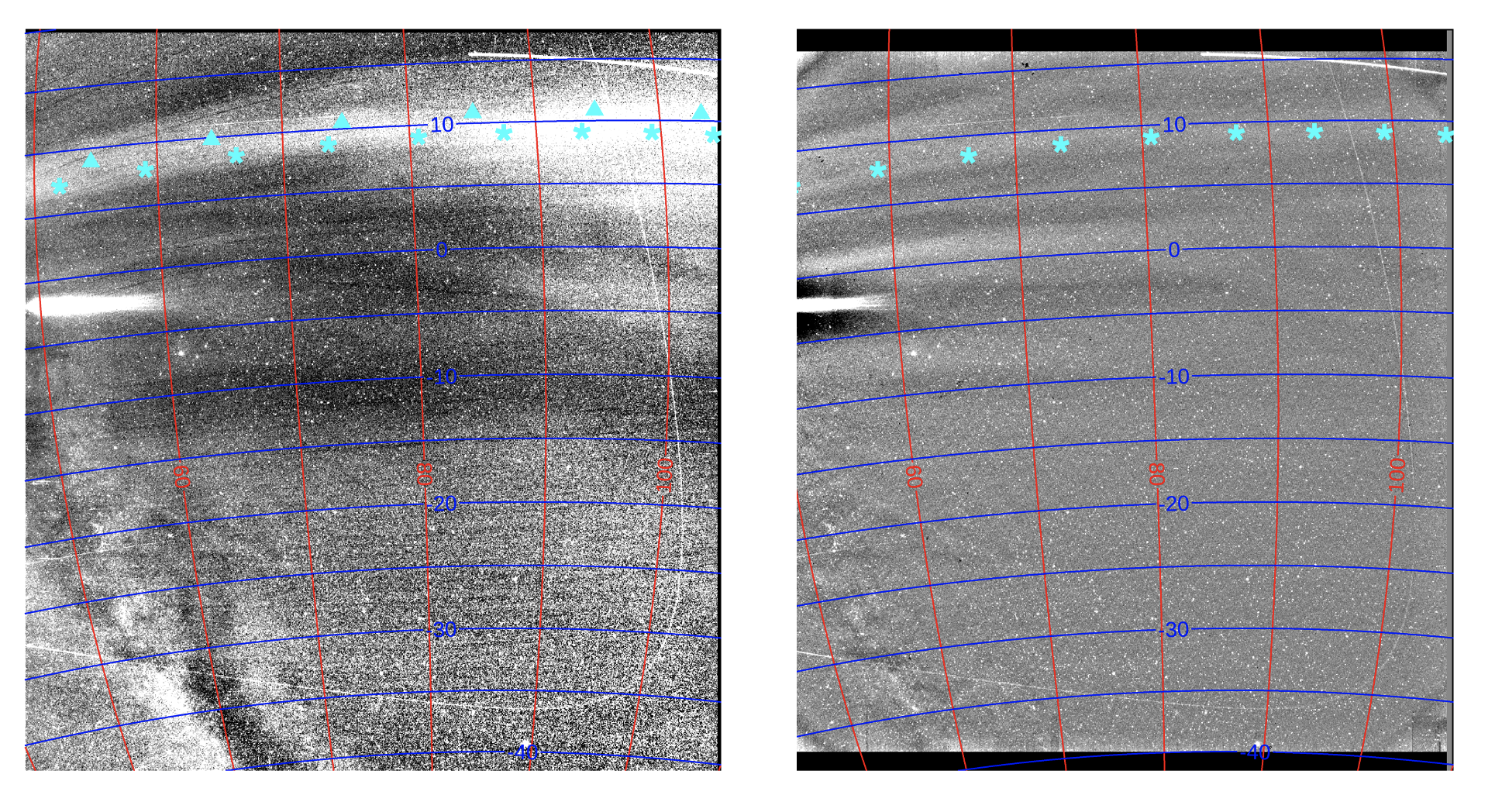}
              }
	\caption{ Left: HCI and Carrington frame solutions for WIS-O High Ray 2 projected onto the WISPR-O LW-processed image for 2021-01-17T09; this is one of the images used in the tracking. Again, the HCI solution is plotted as asterisks and the Carrington solution at triangles. Right: HCI solution projected onto the WISPR-O L3-processed image for the same time. The coronal ray is quite visible in the L3 image. The bright blur is an artifact caused by Venus. }
    \label{fig:wisohigh2}
       \end{figure}

On the left in Figure~\ref{fig:wisohigh2}, the two solutions are projected onto one of the images used in tracking, WISPR-O LW image on 2021-01-17T09, again with the HCI solutions as asterisks and the Carrington solutions as triangles. Points for both solutions start at 17 \rsun\  and have a fixed spacing of 1 \rsun.  Here, as in the previous case, both solutions project equally well onto the ray and the $\chi^2$’s are also similar. The uncertainty in the longitude, calculated from the difference in the HCI and Carrington solutions as described above, was smaller here: 5$ ^\circ$. The bright blur at the inner edge is an artifact caused by Venus.

The right image in Figure~\ref{fig:wisohigh2} shows the projection of the point for the HCI solutions  onto an L3-processed image (see discussion of Figure~\ref{fig:polarplot} above). 
Note that, unlike in Figure~\ref{fig:wisohigh1}, the tracked coronal ray is quite visible across the entire FOV of WISPR-O. This image was taken about five hours before the in-situ instrument detected three closely spaced current sheet  crossings between 14 and 15 UT on 2021-01-17 (Bale 2022, private communication), suggesting a folded current sheet. 
The bright coronal ray seen at $0^\circ$ latitude is apparently in the PSP orbit plane since it shows no  apparent change in latitude during the sequence. The ray is probably due to the plasma near the heliospheric current sheet, crossed by PSP  about five hours later. 

\subsection {WIS-I Low Ray }
The angular coordinates for the third feature, WIS-I Low Ray, as determined by the Tracking and Fitting procedure for both Carrington and HCI frames, are given in Table I, along with the other information. The angular coordinates are plotted in Figure~\ref{fig:psporbit} in green, with the HCI solution as the short line and the Carrington solution as a triangle.  The range of longitudes covered by PSP during the tracking of this feature is evident from the range of PSP longitudes during the tracking shown in Figure~\ref{fig:psporbit}  as the green bar on the x-axis. 

\begin{figure}    
\centerline{\includegraphics[width=1.0\textwidth,clip=]{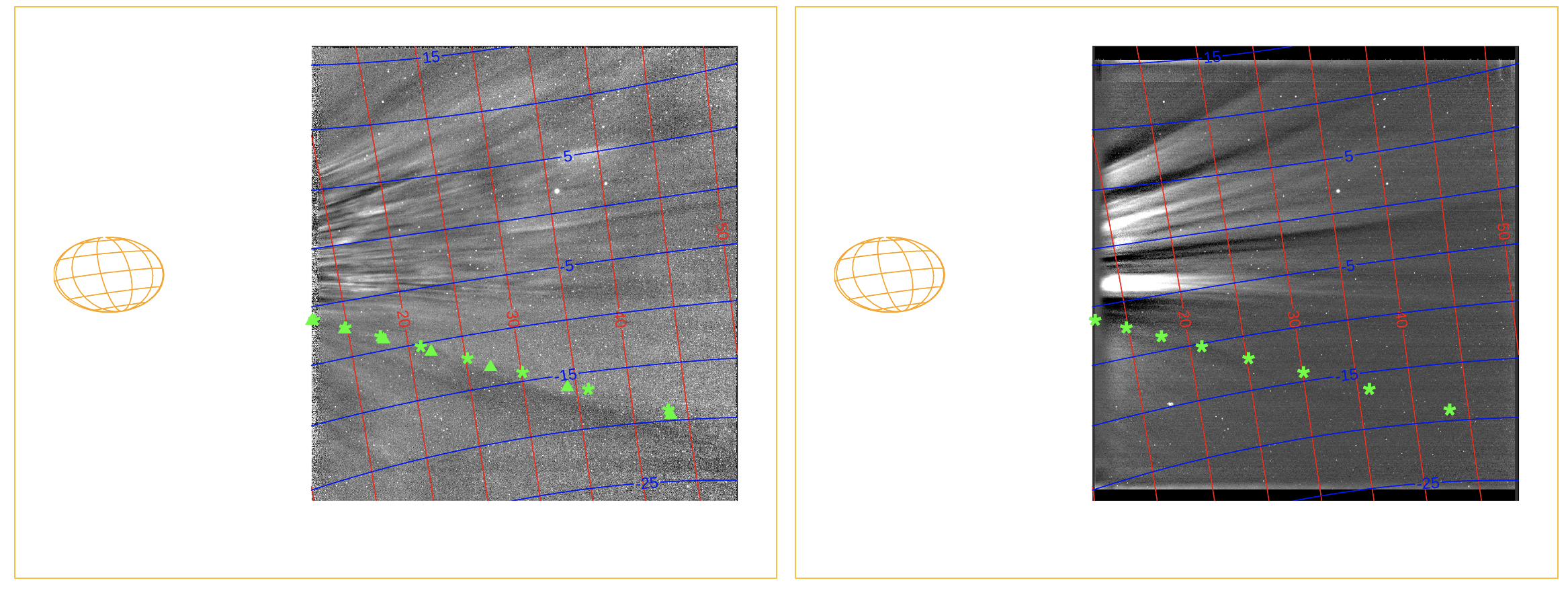}
              }
	\caption{ Left: HCI and Carington frame solutions for WIS-I Low Ray projected on the WISPR-I LW-processed image for Janurary 18 at 01 UT; this is one of the images used in the tracking. Again, the HCI solution is plotted as asterisks and the Carrington solution at triangles. Right: HCI solution projected onto the WISPR-I L3-processed image at the same time. The ray is barely visible in the L3 image. 	}
 \label{fig:wisilow}
    \end{figure}

On the left in Figure~\ref{fig:wisilow}, the two solutions are projected onto one of the images used in tracking, the WISPR-I (13.5$^\circ$ - 53$^\circ$ elongation from Sun center) LW image at 2021-01-18T03, again with the HCI solutions as asterisks and the Carrington solutions as triangles. The points for the HCI solution start at 10 \rsun, the points for the Carrington solution start at 12 \rsun, both with fixed spacing of 1 \rsun. Here, as in the previous cases, both solutions project equally well onto the image and the $\chi^2$’s are also similar (Table 1). The uncertainty in the longitude, calculated from the difference in the HCI and Carrington solutions as above, was $6^\circ$. On the right in Figure~\ref{fig:wisilow}, the HCI solution in projected onto an L3-processed image as was done for the first two rays in earlier figures.  Here, the ray is barely visible in the L3 images, and, since the ray is about the same distance, this suggests less density enhancement along the LOS.  In Figure~\ref{fig:wisilow}, the yellow globe of the Sun is shown to scale in size and distance. The image is projected in the WISPR-I camera frame, which is only valid within its FOV, causing the  distortion of the Sun globe.
Both solutions for this ray segment are also plotted in Figure~\ref{fig:psporbit}, using the same assigned colors and symbols. 
   
\subsection{WIS-O Low Ray}  
\begin{figure}    
\centerline{\includegraphics[width=1.0\textwidth,clip=]{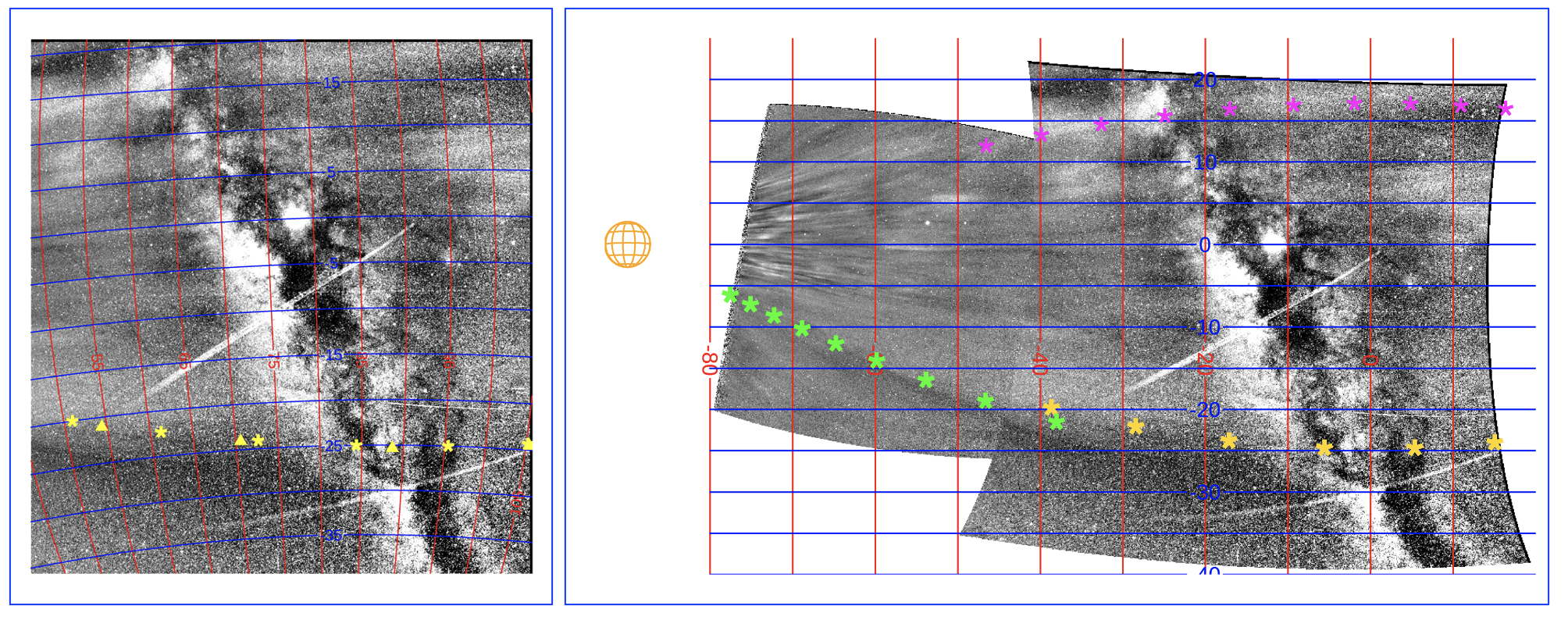}
              }
	\caption{Left: HCI and Carrington frame solutions for WIS-O Low Ray (yellow) projected on the WISPR-O LW-processed image for 2021-01-18T03. Again, the HCI solution is plotted as asterisks and the Carrington solution as triangles. Right: HCI solutions for both the WIS-I (green) and WIS-O (yellow) projected onto a combined WISPR inner and outer image. For reference, we have also projected the WIS-O High Ray 1 HCI solution (magenta)on the image. The combined image is in the PSP orbit frame in which  $0^\circ$ latitude is the PSP orbit frame. The Milky Way is quite evident in this image.
	}
  \label{fig:wisolow}
     \end{figure}

The angular coordinates for the fourth feature, WIS-O Low Ray, as determined by the Tracking and Fitting procedure for both Carrington and HCI frames, are given in Table I, along with the other information. The angular coordinates are plotted in Figure~\ref{fig:psporbit} in yellow, with the HCI solution as the short line and the Carrington solution as a triangle.  The range of longitudes covered by PSP during the tracking of this feature is  shown in Figure~\ref{fig:psporbit} as the {\bf thin yellow line just above}  the x-axis. On the left in Figure~\ref{fig:wisolow}, the two solutions are projected onto one of the WISPR-O images used in tracking, the LW image at 2021-01-18T03, again with the HCI solutions as asterisks and the Carrington solutions as triangles. The points for both solutions start at  17 \rsun\ with a fixed spacing of 1 \rsun. Here, as in the previous cases, both solutions project equally well onto the image and the $\chi^2$’s are also similar (Table 1). The uncertainty in the longitude, calculated from the difference in the HCI and Carrington solutions as above, was $5 ^\circ$.

Note that the coordinates are nearly the same as the  WIS-I Low Ray and, thus, we conclude these are just independent trackings of the same ray. This is confirmed by the image on the right in Figure~\ref{fig:wisolow}, a composite image of the WISPR-I and WISPR-O images for 2021-01-18T03. The ray can be clearly seen to be continuous across the entire WISPR FOV.
Over this composite image, we have plotted both the WIS-I
(green) and WIS-O (yellow) Low Ray HCI solutions  for their respective radial ranges.
The projection here is in the PSP orbit frame, defined in Section 2.1, so $0^\circ$ latitude is the PSP orbit plane. The Milky Way is quite evident in both images. Both solutions for this ray segment are also plotted in 
Figure ~\ref{fig:psporbit}, using the usual assigned colors and symbols, again indicating WIS-I and WIS-O Low Ray are segment of the same coronal ray. The combined radial extent of this ray is about   12 \rsun.  
[In the plot in Figure ~\ref{fig:psporbit}, the yellow WIS-O Low Ray HCI solution $(102.5^\circ)$ is somewhat obscured by the magenta Carrington WIS-O High Ray 2 $(93.7^\circ)$.]


\section{Comparison with Images from Other Coronagraphs and with Synoptic MHD results}

In this section, we compare the locations determined by the Tracking and Fitting method with simultaneous observations from two coronagraphs, LASCO/C3 and STEREO-A/COR2 to validate our results and with synoptic results from the CORHEL MHD model.

 \begin{figure}    
\centerline{\includegraphics[width=1.0\textwidth,clip=]{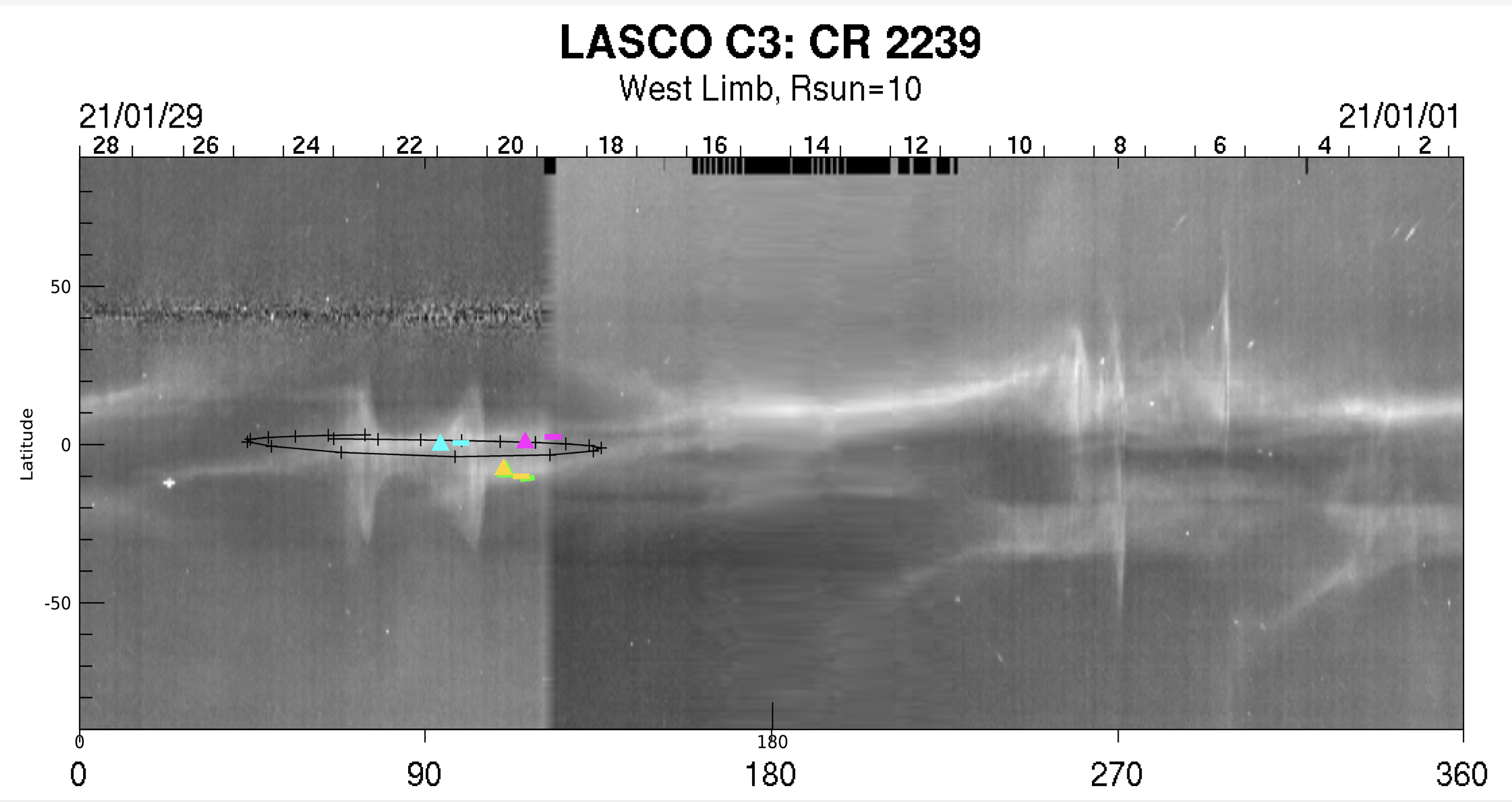}
              }
	\caption{ Comparison of results for coronal ray locations with LASCO/C3 synoptic map for Carrington rotation 2239 for 10 \rsun. The overlay of the locations and the PSP orbit is the same as that in Figure~\ref{fig:psporbit} which explains the colors and symbols.  All the ray locations found by the Tracking and Fitting technique fall within the bright features in the synoptic image, giving us  confidence in our results. 	}
\label{fig:c3synoptic}  
   \end{figure}
  
In Figure~\ref{fig:c3synoptic},  the solutions for  the coronal ray locations for both HCI and Carrington solutions are plotted on a LASCO/C3 west limb synoptic map at 10 \rsun \ for Carrington rotation 2239. LASCO synoptic maps and information is available at https://lasco-www.nrl.navy.mil/. 
The synoptic C3 maps are created in a manner analogous to creating synoptic magnetograms, but here the data at a fixed radius on the west limb in the coronagraph image is collected for each day, converted to a strip and the strips stacked in time/Carrington longitude to create the Carrington synoptic map.
As with synoptic magnetograms, time (shown on the upper x-axis) goes from right to left. 
The brightest band in C3 synoptic maps is generally interpreted as  showing the location of the sheet of plasma encasing the  HCS, as discussed in the Introduction. 
The overlay of the ray positions uses the same plot of coronal ray positions relative to the PSP orbit as used in Figure~\ref{fig:psporbit} and  the color coding and the symbols are unchanged. The locations of both solutions for all four features  fall on the bands of brightness seen by LASCO/C3, indicating that coronal rays were present at those latitudes and longitudes when viewed by LASCO at the time on the top axis. This gives us some confidence that our technique provides accurate 3D locations since the view from SOHO was about $40^\circ$ from that of PSP. Unlike the synoptic maps of the HCS made from models, such as the various PFSS models, which use static synoptic magnetograms, the synoptic maps made from coronagraph data show time variation as the rays move up or down in response to coronal restructuring, streamer blowouts and CMEs. 

\begin{figure}    
\centerline{\includegraphics[width=1.0\textwidth,clip=]{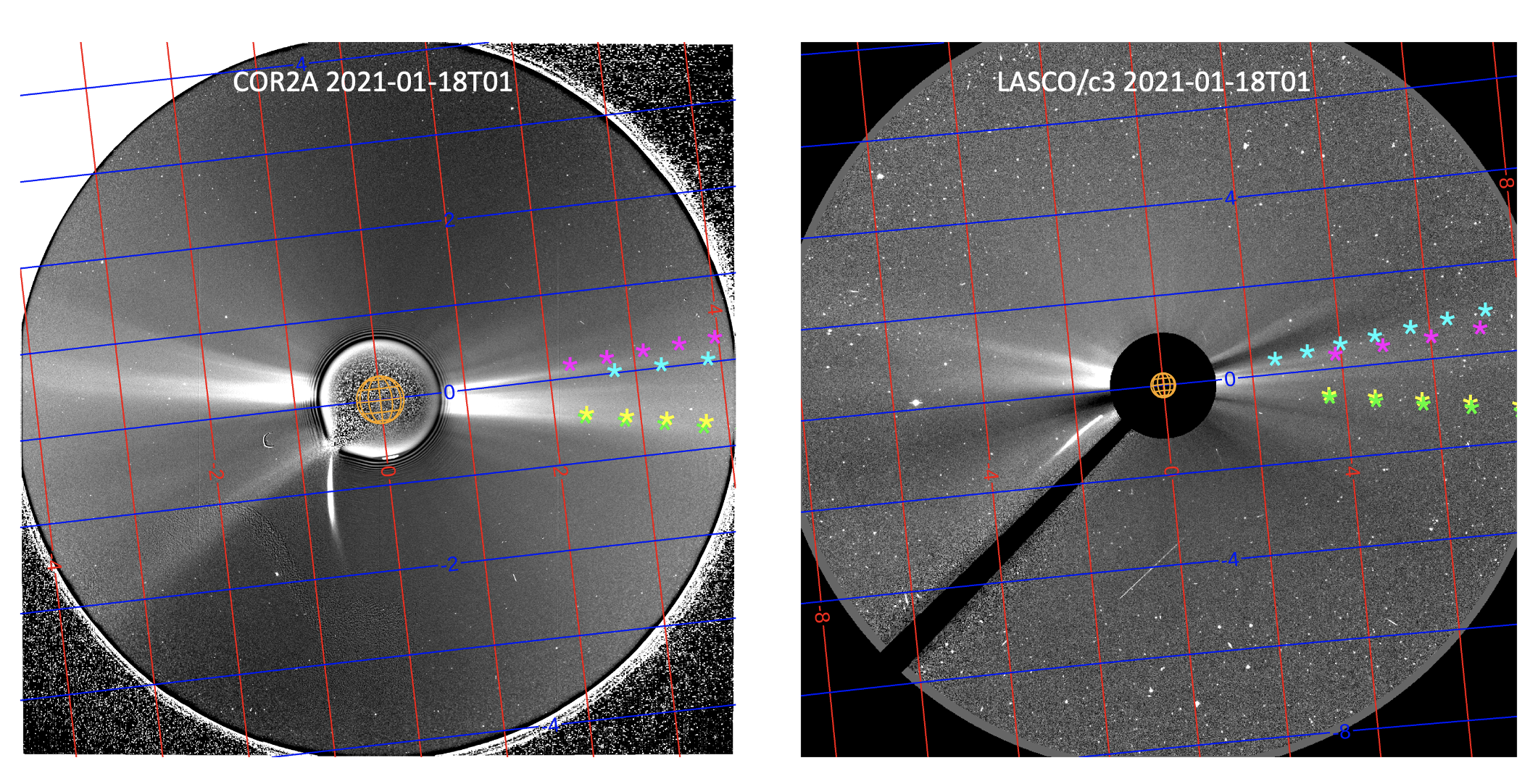}
              }
	\caption{Left: Projection of the four HCI coronal ray solutions onto a COR2A image at 2021-01-18T01. The rays fall on or near rays seen from STA even though the view points are well separated 
	(Figure~\ref{fig:polarplot}). 
	Right: Projection of the four HCI solutions on to a LASCO/C3 images at the same time.  The bright curved streak is a comet tail.	}
\label{fig:projcor2c3}  
   \end{figure}

More detailed comparisons were made with single images from LASCO/C3 and COR2A. 
In  Figure~\ref{fig:projcor2c3}, 
the coronal ray HCI solutions for all four ray segments are projected onto COR2A (left) and LASCO/C3 (right) images from  2021-01-18T01. The coronal ray solutions are plotted in their assigned colors.  The points projected onto the COR2A image start at 10 \rsun\  and are separated by 2 \rsun. The points projected onto the LASCO/C3  image start at 15 \rsun\  and are separated by 4 \rsun. The bright curved feature is a comet tail. 
Note that the radial extent of the segments shown in Figure~\ref{fig:projcor2c3} do  not correspond to the radial extent seen and tracked in the WISPR images (only WIS-I Low Ray extended into 10 \rsun).  The radial extent of the tracked segments were given in Table 1 and illustrated in the plot in Figure~\ref{fig:polarplot}.
Earth and STEREO-A were separated by about 56$^\circ$, {\bf as shown in Figure~\ref{fig:polarplot}}. The rays were somewhat closer to COR2A’s plane-of-the-sky than to LASCO’s, which may explain why they are brighter in COR2A. The LASCO/C3 image is one of those used to make the synoptic Carrington map in Figure~\ref{fig:c3synoptic}. The magenta, yellow and green rays were all visible to WISPR at this time; the cyan ray was seen about 11 hours earlier (see Table I). We can project it on this image because the HCI angles do not change in time. The good agreement with the rays seen in the COR2A and LASCO/C3 images increase our confidence in the technique, since we expect the coronal rays to fall within the coronal plasma sheets observed by these coronagraphs.

\begin{figure}    
\centerline{\includegraphics[width=1.0\textwidth,clip=]{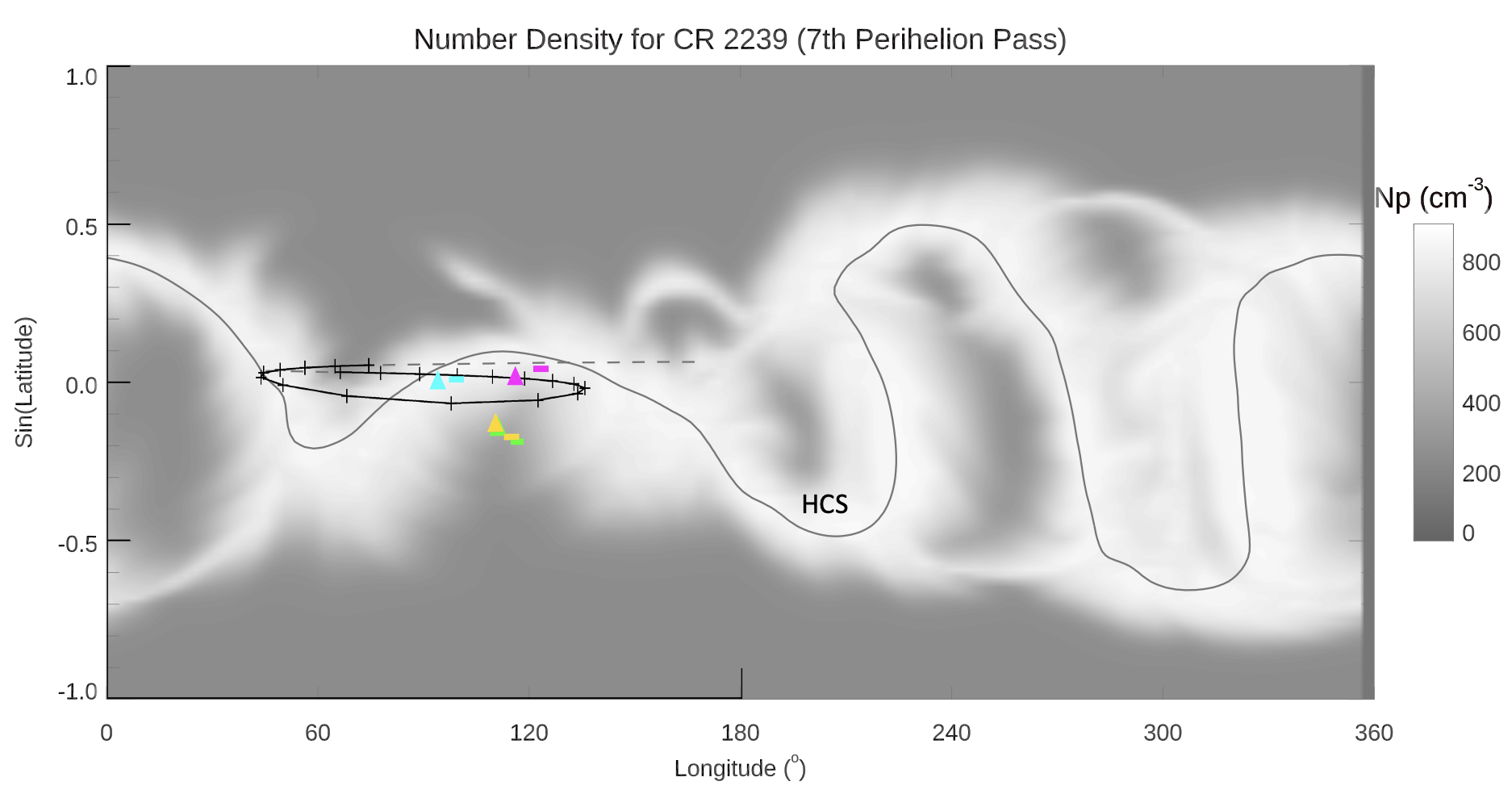}
              }
	\caption{ Comparison of the extracted coronal ray locations with results from the 3D MHD model CORHEL. A plot of the  ray locations relative to the PSP orbit as shown in 
in Figure~ \ref{fig:psporbit}, with no change in the symbols, has been laid over a contour plot of plasma density and  HCS from the model (see text).
	}
\label{MASsynoptic}  
   \end{figure}

Next, in Figure~\ref{MASsynoptic}, we compare the Carrington solution angles of the  coronal rays tracked
to the location of the heliospheric current sheet as computed by an MHD model 
for Carrington rotation 2239. The same plot of the  ray locations relative to the PSP orbit 
in Figure~ \ref{fig:psporbit}, with no change in the symbols, has been overlayed on  a contour plot of plasma density with the computed HCS indicated.
As discussed in the Introduction,  the first explanation of coronal rays at solar minimum \citep{Wang1997} was that the rays result from folds in a thin plasma sheet surrounding the HCS, and, thus, it is of interest to compare the location of coronal rays as determined by our Tracking and Fitting technique to the  predicted location of the HCS. 
In this study, we use results from the 
CORona-HELiosphere (CORHEL) model suite. The modeling region is separated into coronal and heliospheric domains, with the boundary typically lying at $30 R_S$.
A static, synoptic magnetogram provides the boundary condition at the inner-most boundary and the models are run until a steady state is reached.
We use thermodynamic solutions, where energy transport processes are considered, albeit in a semi-empirical way. While these results accurately capture the structure of the coronal magnetic field, they do not reproduce accurate speed and density variations. To address this, rather than using the plasma parameters produced directly from the coronal model to drive the heliospheric model, we employ the ``Distance from the Coronal Hole Boundary'' (DCHB) to derive the speed profile at 30 \rsun, which, has been shown to provide velocity maps that better match in situ measurements than the first-principles results \citep{riley01a}. Density values are then estimated by assuming momentum conservation. Pressure (or temperature) is derived by assuming transverse pressure balance. This model approach (and other variants) are described in more detail by \citet{riley21e}, and references therein. Finally, the location of the heliospheric current sheet (HCS) is inferred from contours where $B_r = 0$.

The comparison in Figure~\ref{MASsynoptic} shows that while the location of the coronal rays always fall in or near the band of denser solar wind predicted by the model, only  WIS-O High Rays 1 and 2 fall near the HCS predicted by the model. 
This suggests these rays may be the result of folds in the HCS or density variations along the sheet of the plasma surrounding the HCS \citep{Thernisien2006}.
The low ray is quite far from the predicted HCS suggesting a different source such as a pseudo-streamer. Alternatively, the model may be inaccurate, since  it cannot capture dynamic changes in the corona using a static magnetic boundary condition.
Note that the complex structure of the white light bands seen in the LASCO/C3 synoptic Carrington map, as well as the triple HCS crossing observed by PSP in situ instruments,  would not be explained by the predictions of the model.

\section{Summary and Discussion}

In this paper we have described and implemented a technique for determining the 3D location 
of coronal ray segments observed in a sequence of images from the heliospheric imager WISPR on the {\it Parker Solar Probe}.  The basis for the technique is the 
super-rotational angular speed of PSP near perihelion. As a result of the speed, WISPR observes
coronal rays as they approach and pass over or under the spacecraft. The apparent change in  latitude of a ray as approached by PSP, together with the assumption that the ray segments lie along radial lines in a heliocentric reference frame, is sufficient information to determine the angular coordinates
of the rays. Solutions can be found assuming a radial line in the HCI frame (fixed ray in inertial space) or assuming a radial line in the Carrington frame (rigid rotation with the Sun). 
The technique is a modification of the Tracking and Fitting used to determine the trajectories of CMEs and "blobs" previously \citep{Liewer2020}.  

We presented the fitting results for the coordinate determination of four sets of tracking data. This represented three coronal rays because two of the data sets corresponded to the tracking of the same ray, one in WISPR-I images and one in WISPR-O images.  For the three coronal rays, solutions were found in both the HCI and Carrington frames. Both the HCI frame and Carrington frame solutions fit the data equally well, based on the fitting statistics, and, as shown above by comparisons of the two solutions with each other and with the WISPR images. We opted to use the difference in the two solutions to estimate an overall error in the location determination. We compared the location of the three rays with observations from other white light imagers with different viewpoints, SOHO/LASCO and STEREO/COR2A, and found that the three rays fell within the regions of coronal rays seen by these coronagraphs, verifying the technique. The comparison also indicates that we are seeing finer scale coronal rays that are unresolved from 1 AU.  The determination of the locations of coronal rays using the technique presented can be used
in the future to validate higher resolution and dynamic MHD models that model the evolution of the corona.

We have made a limited attempt to understand the origin of the three rays analyzed.  Two of the rays, High Rays 1 and 2 were very close to PSP orbit plane  and were observed close to the time that the in situ data showed multiple HCS crossings. In addition, a comparison with the HCS predicted by an MHD model showed that these rays were close to the HCS. Thus, these probably resulted from folds or inhomogeneities in the plasma sheet around the HCS. The other ray, WIS-I and WIS-O Low Ray, was further from the predicted HCS, which could be associated with plasma from a pseudo-streamer or may reflect the effect of flux emergence or coronal restructuring not captured in the MHD model using a static magnetic boundary condition. 
This ray did fall on a bright band in the LASCO/C3 synoptic map (Figure~\ref{fig:c3synoptic}), giving confidence to our location determination. The synoptic  map showed two separated bright bands at the longitude of this ray. Synoptic coronagraph maps show temporal variations in the coronal which  can not be captured using a static synoptic magnetogram as the boundary condition as in the  MHD model.

Note in the last three frames of Figure~\ref{fig:2rays}, the two marked rays, High Outer Ray 1 (magenta) and Low Outer Ray (yellow), appear to enclose somewhat brighter region of the image.
 The brighter wedge is also apparent in both WISPR-I images in Figure~\ref{fig:wisilow}, where we see that
the brighter wedge in the LW images on the left corresponds to the region containing the bright coronal rays in the L3 (calibrated) image on the right, suggesting that this wedge corresponds to the denser plasma of the streamer belt.
Note that the LW processing reveals that there are very small-scale density fluctuations along the rays. Comparison of the left and right images shows that all the bright coronal rays in the L3 image contain these very small-scale density fluctuations, indicating that plasma flowing out is highly variable on very small scales, as well as on the larger scales of rays, blobs and CMEs. This is consistent with the observations in \citet{DeForest2018}. 
The resolved smaller rays with small scale density fluctuations are better seen in subsequent orbits with a smaller perihelion \citep[][in press]{Howard2022}. 

The bright wedge containing the brightest streamers is also apparent in the right image of Figure~\ref{fig:wisolow} which shows the full WISPR FOV image with the brighter region bracketed by High Outer Ray 1 (magenta) on the top and, on the bottom, by the Low Outer Ray (yellow) and its extension into the inner FOV Low Inner Ray (green). If we assume that the upper and lower tracked rays, marked in Figures~\ref{fig:2rays} and \ref{fig:wisolow}, enclose the dense plasma region, we can estimate the thickness in latitude of this region from our knowledge of the latitude of these two rays. Subtracting the latitudes of the High Outer Ray 1 (magenta) and Low Outer Ray (yellow) for both the Carrington and HCI solutions (Table 1)  gives us an estimated width of approximately $7^\circ$ - $12^\circ$, much smaller than the {\it apparent} latitudinal width seen in the images.  The wider apparent width is due to WISPR’s view of the streamer belt from within.  

\begin{acks}
We thank Russ Howard for his support and for many fruitful conversations throughout this work.
We also gratefully acknowledge the help and support of Brendan Gallagher, Phil Hess, Nathan Rich, and the rest of the WISPR team. We thank the reviewer for the insight on the bias of the Carrington solutions towards yielding longitudes closer to PSP than the HCI solutions, which is now discussed in the text. {\it Parker Solar Probe} was designed, built, and is now operated by the Johns Hopkins Applied Physics Laboratory as part of NASA's Living with a Star (LWS) program (contract NNN06AA01C). The Wide-Field Imager for Parker Solar Probe (WISPR) instrument was designed, built, and is now operated by the US Naval Research Laboratory in collaboration with Johns Hopkins University/Applied Physics Laboratory, California Institute of Technology/Jet Propulsion Laboratory, University of Goettingen, Germany, Centre Spatiale de Liege, Belgium and University of Toulouse/Research Institute in Astrophysics and Planetology. WISPR data is available at wispr.nrl.navy.mil/wisprdata.
The work of P.~C.\,Liewer, J.~R.\,Hall, and P.\,Penteado was conducted at the 
Jet Propulsion Laboratory, California Institute of Technology under a contract from NASA. 
J.\,Qiu is partly supported by NASA's HGI program (80NSSC18K0622).F. Ark was supported by the NSF Research Experiences for Undergraduates (REU) program at Montana State University (1851822). A.\,Vourlidas  and G. Stenborg are
supported by the WISPR Phase E program at APL.  P. Riley gratefully acknowledges support from NASA (80NSSC18K0100, NNX16AG86G, 80NSSC18K1129, 80NSSC18K0101, 80NSSC20K1285, and NNN06AA01C).

\end{acks}

\section*{Disclosure of Potential Conflicts of Interest}
The authors declare that they have no conflicts of interest.

\appendix


To more precisely determine the position of a feature in a solar equatorial frame, either the HCI frame or Carrington frame, from measurements of the features position in the PSP orbit frame (defined in Section 2), we take into account the inclination $\epsilon$ of PSP's orbit with respect to the solar equator, and properly transform the feature's position between the solar equatorial and PSP  orbit frames. The details of the transformation and correct equation relating the feature's longitude $\phi_2$ and latitude $\delta_2$ in a solar equatorial frame (HCI or Carrington frame) to the [$\gamma$, $\beta$] measurements in the PSP orbit frame (see Figure~\ref{fig:geom}) are given in the Appendix to Article \citetalias{Liewer2019}.  With the transformation, we re-calculate the ratio $\mathcal{R} \equiv \tan\beta/\sin\gamma$, and then expand the $\mathcal{R}$ equation to only include up to the first order terms of $\epsilon$, as given below.
\begin{flalign}
\mathcal{R} \equiv \frac{\tan\beta}{\sin\gamma} = \frac{\tan\delta_2}{\sin(\phi_2 - \phi_1)} \left(1 - F\sin\epsilon\right), & \label{eq:correct}
\end{flalign}
with 
\begin{flalign}
 F(\phi_2,\delta_2, \phi_1) \equiv \frac{\sin\phi_2}{\tan\delta_2} + \frac{\tan\delta_2\cos\phi_1}{\sin(\phi_2 - \phi_1)}. & \label{eq:F}
\end{flalign}

PSP's orbit intersects with the solar equator at PSP's ascending node, which is chosen as the $x$ axis in both frames, and $\phi_2$ and $\phi_1$ used in Equation~\ref{eq:correct} are measured with respect to this axis. Therefore, $\phi_2$ in Equation~\ref{eq:correct} is offset from the HCI or Carrington longitude by a constant, which is the HCI or Carrington longitude of the ascending node of PSP's orbit, and this constant is determined from the ephemeris. When $\phi_2$ is determined from the fitting procedure, the offset will be corrected to uncover the feature's HCI or Carrington longitude.

With the assumption that a corona ray segment lies along a radial line extending from the solar center in either the HCI frame or Carrington frame (i.e. rigid rotation with the Sun), all features along the same ray segment have the same $\phi_2$ and $\delta_2$ in a given image at a given time.  Therefore, in principle, the ratio $\mathcal{R} \equiv (\tan\beta/\sin\gamma)$ of all these features is also a constant in a given image. With this notion, in each image, we measure [$\gamma$, $\beta$] at a few locations along the same ray segment, and derive the mean ratio $\langle\mathcal{R}\rangle$ and its deviation $\Delta \mathcal{R}$ for each time (each image in the sequence). We then use a curve fitting procedure provided in the Solar Software (SSW) to conduct the Levenberg--Marquardt least-squares fit of the time sequence of measured $\langle\mathcal{R}\rangle$ to Equations~\ref{eq:correct},  using $\Delta \mathcal{R}$ as the measurement uncertainty so that the least-squares fit is weighted by this uncertainty. The fit returns $\phi_2$ and $\delta_2$, the uncertainties in these fitting parameters, and the reduced $\chi^2 $, defined as 
\begin{equation}
\chi^2 = \frac{\sum{\frac{(\mathcal{R}_m - \langle\mathcal{R}\rangle)^2}{(\Delta\mathcal{R})^2}}}{\nu}, \label{eq:chisq}
\end{equation}
where $\mathcal{R}_m$ is the ratio $\mathcal{R}$ calculated from the solution, and $\nu$ is the number of the degrees of the freedom in the fitting, equal to the number of images minus the number of parameters to fit).
We consider a fit as a good fit if the reduced $\chi^2$ is around unity. Note that in some examples in Table~\ref{table:summary}, the reduced $\chi^2$ becomes substantially smaller than unity; this may indicate that $\Delta \mathcal{R}$ is an over-estimate of the true measurement uncertainty, which is hard to determine accurately.

The convergence of a non-linear least-squares fit often depends on the initial input. We calculate the zeroth-order solutions (i.e., assuming $\epsilon = 0$) of a feature's position $[\phi_2, \delta_2]$  from Equation~\ref{eq:1} and use them as the initial input for the fit to the corrected Equation~\ref{eq:correct}. The steps to estimate the initial guess are given in detail in Article \citetalias{Liewer2019}. 

\begin{figure}    
\centerline{\includegraphics[width=1.0\textwidth,clip=]{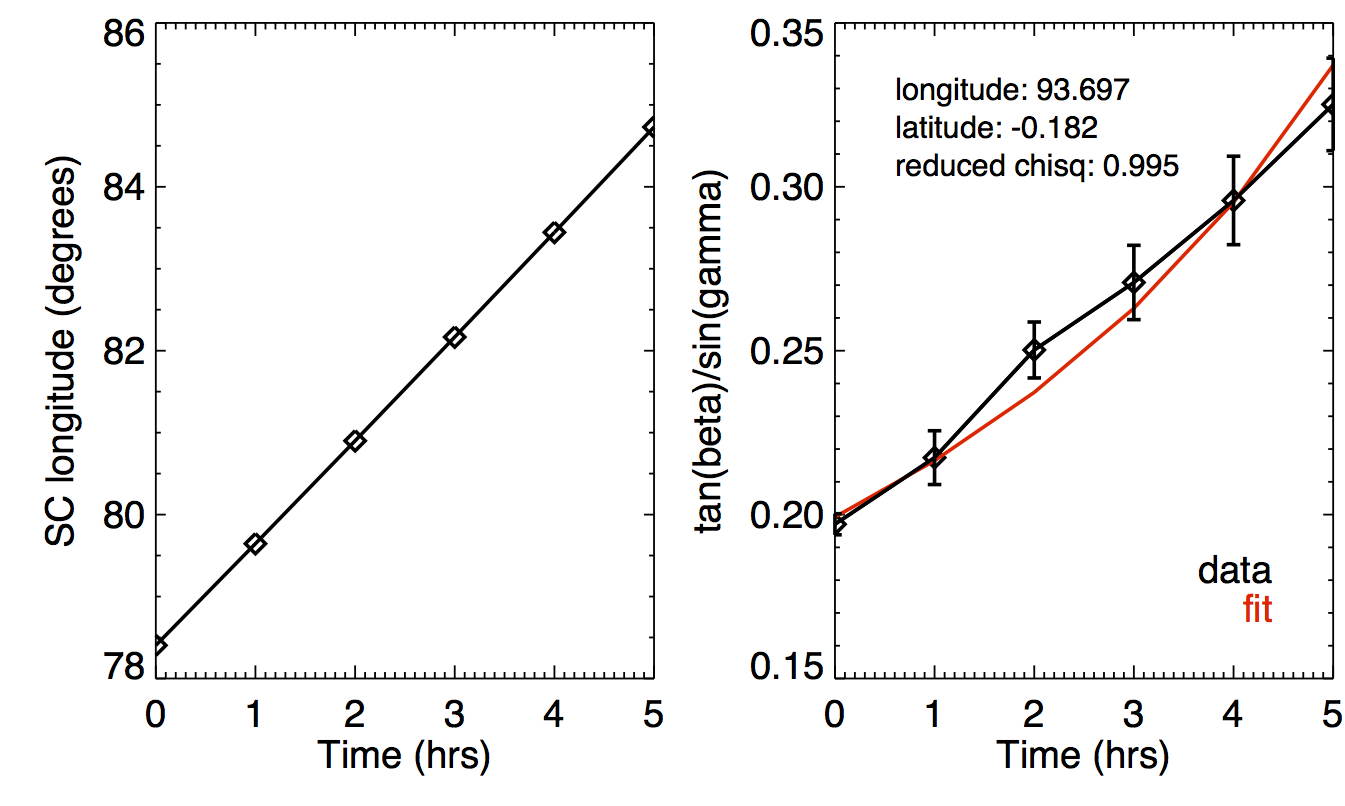}
              }
	\caption{An example of finding the solution of a ray's Carrington longitude and latitude from fitting. Left: Carrington longitude of PSP during the 5-hr observation of the ray segment. Right: the measured mean ratio $\langle\mathcal{R}\rangle$ (black) and the fit (red). Vertical bars show the measurement uncertainty $\Delta \mathcal{R}$ derived from $[\gamma, \beta]$ measurements at multiple locations along the ray segment.}
\label{fig:fit}  
   \end{figure}

Finally, we note that the corrected fitting equation, Equation~\ref{eq:correct}, takes into account small angles, so it provides reasonable solutions for small $\delta_2$ when the feature is nearly located in the solar equatorial plane but not in PSP's orbit plane. In this case, the lead term in the $\mathcal{R}$ expression in Equation~\ref{eq:correct} becomes $$ \mathcal{R} \approx \frac{\tan\delta_2 - \sin\phi_2\sin\epsilon}{\sin(\phi_2 - \phi_1)}. $$ 
Seen in Figure~\ref{fig:fit}, for a ray lying nearly in the solar equatorial plane $\delta_2 \approx 0$, the $\beta$ angle out of PSP's orbit plane becomes substantial as PSP moves very close to the ray, and $\langle\mathcal{R}\rangle$ grows rapidly, allowing for a reasonable solution from the fit.

\newpage

\bibliographystyle{spr-mp-sola}
\bibliography{psp}


\newpage

\begin{table}
\caption{Summary of Solution Results}
\label{table:summary}
\begin{tabular}{lcccccc r@{.}l c} 
  \hline
{\bf Coronal Ray} & WIS-O High Ray 1 &WIS-O High Ray 2 & WIS-I Low Ray & WIS-O Low Ray  \\
  \hline
  Plotting color & magenta & cyan & green & yellow \\
  tracking start time & 2021-01-17T22:05   & 2021-01-17 T09:05 &  2021-01-18T00:00& 2021-01-18T01:05 \\
  tracking end time & 2021-01-18T03:05   & 2021-01-17 T14:05 &  2021-01-18T04:00. & 2021-01-18T06:05 \\
  \hline
  {\bf CAR frame } & & & & \\
  longitude ($^\circ$) & 115.70$\pm$1.35 & 93.69$\pm$  0.82 &	110.70$\pm$0.67	&110.10$\pm$0.15\\
  latitude ($^\circ$) & 0.59$\pm$0.32 & -0.18$\pm$0.15&	-8.60$\pm$0.36&	-7.90$\pm$0.09\\
  reduced $\chi^2$ & 0.13 & 0.99&	0.04&	0.84 \\
  Radial extent (\rsun) & 16-24 & 17-23 & 10-17  &	18-22\\
  {\bf HCI frame } & & & & \\
  longitude ($^\circ$) & 109.00$\pm$1.78 & 77.40$\pm$1.14 &	103.20$\pm$1.06 &	102.50$\pm$0.25 \\
  latitude ($^\circ$) &2.45$\pm$0.46 & 0.52$\pm$0.24&	-10.85$\pm$0.48 &	-10.00$\pm$0.13 \\
  reduced $\chi^2$ & 0.12 & 1.02 &	0.05 &	0.80 \\
Radial extent & 16-28 & 17-25 & 9-17  &	18-23\\
  \hline
  {\bf Error in location} & & & & \\
  error in longitude ($^\circ$) & 7.3 &	5.4	&5.6&	4.6\\
  error in latitude ($^\circ$) & 1.9  &	0.7	& -2.3&	-2.1 \\
  \hline
\end{tabular}
\end{table}

\end{article}

\end{document}